\documentclass[11pt,oneside,letterpaper]{article}
\usepackage{amssymb}
\usepackage{amsmath,bm}
\usepackage[dvips]{graphicx}
\usepackage{setspace}
\usepackage{fancyhdr}
\usepackage[dvipsnames]{xcolor}
\usepackage{ifpdf}
\usepackage{graphicx}
\usepackage{rotating}
\usepackage{comment}
\usepackage[margin=2.5cm]{geometry}
\usepackage[colorlinks=true, linkcolor  = blue,urlcolor=blue,citecolor=violet]{hyperref}
\usepackage{soul}
\usepackage{empheq}
\usepackage{bbold}
\usepackage{graphicx}
\usepackage{float}
\usepackage{bm}
\usepackage{cite}

\def\p{\partial}

\newcommand{\bi}{\begin{itemize}}
\newcommand{\ei}{\end{itemize}}
\newcommand{\bea}{\begin{eqnarray}}
\newcommand{\eea}{\end{eqnarray}}
\newcommand{\be}{\begin{equation}}
\newcommand{\ee}{\end{equation}}

\numberwithin{equation}{section}
\allowdisplaybreaks  

\begin{document}

\vspace*{2.5cm}
\begin{center}
{\LARGE {Three-dimensional de Sitter horizon thermodynamics}}
\end{center}

\vskip10mm
\begin{center}
{\small{Dionysios Anninos and Eleanor Harris}}
\end{center}

\vskip 5mm
\begin{center}
{\footnotesize{Department of Mathematics, King's College London, the Strand, London WC2R 2LS, U.K.}}
\end{center}
\vskip 5mm
\begin{center}
	{\footnotesize{dionysios.anninos@kcl.ac.uk, \quad eleanor.k.harris@kcl.ac.uk}}
\end{center}

\vspace{4mm}
 
\vspace*{0.6cm}

\vspace*{1.5cm}
\begin{abstract}
\noindent
We explore thermodynamic contributions to the three-dimensional de Sitter horizon originating from metric and Chern-Simons gauge field fluctuations. In Euclidean signature these are computed by the partition function of gravity coupled to matter semi-classically expanded about the round three-sphere saddle. We investigate a corresponding  Lorentzian picture --- drawing inspiration from the topological entanglement entropy literature --- in the form of an edge-mode theory residing at the de Sitter horizon. We extend the discussion to three-dimensional gravity with positive cosmological constant, viewed (semi-classically) as a complexified Chern-Simons theory. The putative gravitational edge-mode theory is a complexified version of the chiral Wess-Zumino-Witten model associated to the edge-modes of ordinary Chern-Simons theory. We introduce and solve a family of complexified Abelian Chern-Simons theories as a way to elucidate some of the more salient features of the gravitational edge-mode theories. We comment on the relation to the AdS$_4$/CFT$_3$ correspondence.

\end{abstract}

\newpage
\setcounter{page}{1}
\pagenumbering{arabic}

\tableofcontents

\onehalfspacing

\section{Introduction}

This paper explores properties of massless quantum fields on a three-dimensional de Sitter background. Our main focus is on their contribution to the thermodynamic properties of the de Sitter horizon. The static patch of de Sitter is described by the metric
\begin{equation}\label{sp}
\frac{ds^2}{\ell^2} = -dt^2\cos^2 \rho + d\rho^2 + \sin^2\rho \, d\varphi^2~, 
\end{equation}
where $t \in\mathbb{R}$, $0 \leq \rho \leq \pi/2$, $\varphi \sim \varphi + 2\pi$. The de Sitter horizon resides at $\rho = \pi/2$ and has size $2\pi \ell$. Although the de Sitter horizon resembles the black hole horizon to some degree, the two differ in several regards \cite{Anninos:2018svg}. For instance, the de Sitter horizon has no parameter tuning its associated temperature, heavy quantum fields become increasingly insensitive to its presence, and it behaves oppositely to a black hole under absorption of a pulse of energy. As we take the large $\ell$ limit, the de Sitter horizon tends to the null boundary of an empty Minkowski spacetime potentially linking it to the physics of a thermal version of soft modes \cite{Strominger:2017zoo}.\footnote{This limit might also suggest we treat the de Sitter horizon as a boundary \cite{Bousso:1999cb,Banks:2013qpa,Alishahiha:2004md,Compere:2020lrt,Freidel:2021ajp}, although a clear framework for how to do so remains unclear.} As such, a careful and quantitative exploration of the features of the de Sitter horizon is warranted.

We will focus on the relation of the static patch to the partition function over a Euclidean three-dimensional round sphere, which is the Euclidean continuation $t \to -i \xi$ of the static patch (\ref{sp}), with $\xi \sim \xi+2\pi$. For theories of gravity coupled to matter admitting a semiclassical de Sitter solution, it was proposed by Gibbons and Hawking in \cite{Gibbons:1976ue}, and more recently explored in \cite{Anninos:2020hfj,Law:2020cpj,David:2021wrw}, that the de Sitter horizon entropy is computed by the logarithm of the Euclidean gravitational path integral $\mathcal{Z}_{\text{grav}}$ about the  round sphere (Euclidean static patch) saddle point solution. For a theory of pure Einstein gravity with a positive cosmological constant $\Lambda = +1/\ell^2$, and Newton constant $G$ in three spacetime dimensions, one has \cite{Anninos:2020hfj,Carlip:1992wg,Guadagnini:1995wv,Castro:2011xb}
\begin{equation}\label{logZgrav}
\log \mathcal{Z}_{\text{grav}} =   S_{\text{dS}} - 3 \log  S_{\text{dS}} + 5\log 2\pi  + i \varphi_{\text{grav}} + \ldots 
\end{equation}
where $S_{\text{dS}} = \pi \ell/2G \gg 1$ is the tree-level Gibbons-Hawking entropy of the de Sitter horizon, which is parametrically large in the semiclassical limit. Although three-dimensional gravity carries no propagating degrees of freedom, the structure (\ref{logZgrav}) is non-trivial and a microscopic understanding is so far lacking (attempts include \cite{Park:1998qk, Maldacena:1998ih,Banados:1998tb,Govindarajan:2002ry,Dong:2010pm}). Unlike the leading term, the subleading terms in (\ref{logZgrav}) stem from perturbative quantum corrections of the gravitational fluctuations. The logarithmic term is related to the residual $SO(4)$ subgroup of the diffeomorphism group preserved by the three-sphere saddle. The constant part is unambiguous in odd spacetime dimensions as it cannot be absorbed into a local counterterm, while the phase $\varphi_{\text{grav}}$ can arise due to the unboundedness of the conformal mode. In the treatment of \cite{Polchinski:1988ua} one finds that $\varphi_{\text{grav}} = - {5 \pi  }/{2}$ to leading order at large $S_{\text{dS}}$. 

Further corrections to (\ref{logZgrav}) will appear as even inverse powers of $S_{\text{dS}}$. Upon coupling the theory to matter fields, the expansion (\ref{logZgrav}) may receive additional contributions. Consider first matter fields that are parametrically heavy with respect to the de Sitter length $\ell$. Integrating them out results in an effective gravitational theory that will be perturbed, to high accuracy, by a local functional of the metric containing higher derivative terms. Provided parity is preserved, due to the absence of local degrees of freedom in three-dimensional gravity, such a theory can always be brought back to an Einstein theory with a cosmological constant through local field redefinitions of the metric. From this perspective, integrating out massless or light fields, or some more general conformal matter theory, is  interesting in that one can affect the details of (\ref{logZgrav}) which are of a more non-local nature. The two-dimensional version of this problem was recently explored in \cite{Anninos:2021ene,Muhlmann:2021clm}. In what follows we focus on the contribution to (\ref{logZgrav}) from Chern-Simons gauge fields. These are massless gauge theories which are under solid theoretical control while producing various interesting modifications to (\ref{logZgrav}). Further to this, at least semiclassically, three-dimensional gravity can be expressed as a Chern-Simons theory \cite{Achucarro:1987vz,witten19882+}. For de Sitter in Euclidean signature this Chern-Simons theory has gauge group $SU(2)_{\kappa+i
\gamma} \times SU(2)_{\kappa-i\gamma}$ with complexified levels $\kappa \pm  i \gamma$ where $\gamma \equiv \ell/4G \in \mathbb{R}^+$, and $\kappa \in\mathbb{Z}$ is the coupling for a gravitational Chern-Simons type of term. In Lorentzian signature the relevant Chern-Simons theory is one with an $SL(2,\mathbb{C})$ gauge group. Such complexified Chern-Simons theories have been the subject of various interesting works including \cite{Witten:1989ip,Witten:2010cx,Gukov:2016njj,Dimofte:2009yn,Dimofte:2016pua,Vafa:2015euh}. Here we find a natural application of these theories within the realm of three-dimensional de Sitter space. 

The relation between three-dimensional gravity and Chern-Simons theory may permit a more direct link between the Euclidean expression (\ref{logZgrav}) and the Lorentzian picture (\ref{sp}). Of particular interest is whether the logarithmic term and further subleading terms in (\ref{logZgrav}) have a simple Lorentzian interpretation. One hope that such a link exists comes from literature on what is known as topological entanglement entropy \cite{kitaev2006topological,Levin:2006zz}, which is a finite contribution to the vacuum entanglement entropy between disconnected regions of space in Chern-Simons theory or some more general topological field theory. In its simplest form, one quantises Chern-Simons theory on $S^2$ spatial slices and considers the entanglement entropy stemming from splitting the $S^2$ into two disks. It is argued \cite{kitaev2006topological,Levin:2006zz,Fendley:2006gr} that this entanglement entropy is computed by the ${\mathcal{S}_0}^0$ component of the modular $S$-matrix ${\mathcal{S}_m}^n$ of the edge-mode theory living at the boundary of the disk. On the other hand, as was already known from the seminal work \cite{witten1989quantum}, ${\mathcal{S}_0}^0$ computes the partition function of Chern-Simons theory on an $S^3$ topology. 

Our goal will be to express the aforementioned results relating the three-sphere partition function to an edge-mode theory in the language of three-dimensional de Sitter space. We would like to sharpen the general (though somewhat schematic) hypothesis that
\begin{equation}\label{hypothesis}
\mathcal{Z}_{\text{grav}} \overset{?}{=} \lim_{\beta\to0^+}  Z_{\text{edge}}[\beta]~,
\end{equation}
where $Z_{\text{edge}}$ is the thermal partition function of a putative edge-mode theory. The edge-modes are evaluated at parametrically high temperature due to their parametric proximity to the de Sitter horizon. In this paper we develop a collection of tools to assess (\ref{hypothesis}). We provide circumstantial evidence for an edge-mode interpretation for the subleading terms of the semiclassical expansion (\ref{logZgrav}) of $\mathcal{Z}_{\text{grav}}$. The interpretation of the leading term $S_{\text{dS}}$ as an entanglement entropy of an edge-mode theory, or any other microscopic origin, is left open. 

In the first part of our discussion, we take the low energy theory to be three-dimensional general relativity coupled to Chern-Simons gauge theory and consider the contribution to  (\ref{logZgrav}) from the Chern-Simons sector.  In the second part of our discussion, we investigate a Lorentzian interpretation for the purely gravitational expression (\ref{logZgrav}) by viewing the gravitational theory as a complexified Chern-Simons theory. 

\subsection{Outline}

In section \ref{general} we discuss the dS$_3$ geometry and the type of low energy effective theories we will study. In section \ref{css3} we review and place several relevant results from the Chern-Simons literature into the context of three-dimensional de Sitter space. In section \ref{csab} we describe in detail the relation between the edge-mode theory residing at the boundary of a spatial disk to the three-sphere partition function for Abelian Chern-Simons theory. The edge-mode theory introduced in this section serves as the basic template for the latter sections. In section \ref{complexCSL} we discuss an Abelian Chern-Simons theory with complexified $U(1)$ gauge group. This is meant to serve as an Abelian toy model for Lorentzian three-dimensional gravity expressed as a Chern-Simons theory with a complexified $SU(2)$ gauge group. In \ref{complexCS} we introduce an Abelian Chern-Simons theory with complexified level that is meant to serve as a simplified toy model of Euclidean gravity expressed as a Chern-Simons theory with a complexified level. In section \ref{oneloop3dgrav} we briefly discuss how to extend the results of sections \ref{complexCSL} and \ref{complexCS} to the non-Abelian case relevant to general relativity. Finally, in section \ref{adscft} we make some comments on AdS$_4$/CFT$_3$ in the case where AdS$_4$ has a three-dimensional de Sitter boundary. The appendices provide further details on various computations presented in the main text. 

\section{Geometry and general setup}\label{general}

In this section we discuss some basic geometric properties of three-dimensional de Sitter space and the type of low energy effective theories we will consider. 

\subsection{Geometry of dS$_3$}
\begin{figure}[H]
	\centering
	\includegraphics[width=0.33\linewidth]{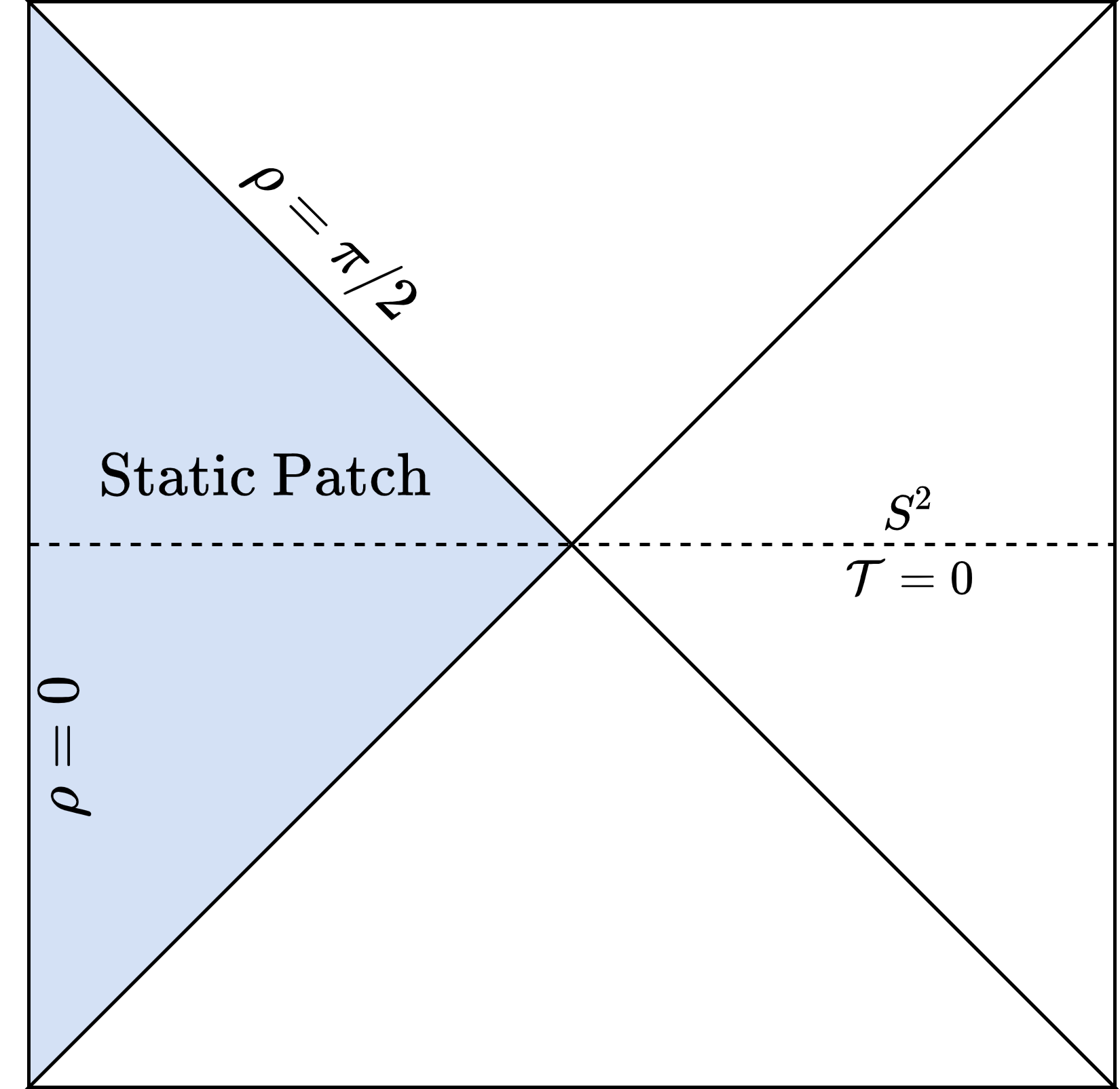}
	\caption{The Penrose diagram of dS$_3$, with the static patch highlighted. Constant time slices are two-spheres, with the $\mathcal{T}=0$ shown explicitly, and the left and right-hand sides of the diagram are the poles of $S^2$. }
	\label{fig:SoftdeSitterPenrose}
\end{figure}

Three-dimensional de Sitter space admits a global chart described by the following metric
\begin{equation}\label{global}
\frac{ds^2}{\ell^2} = -d\mathcal{T}^2+ \cosh^2 \mathcal{T} \, d\Omega^2~, 
\end{equation}
where $\mathcal{T}\in \mathbb{R}$ and $d\Omega$ describes the round metric on the unit two-sphere. The above spacetime is one of constant and positive Ricci scalar, and the length-scale $\ell$ characterises the curvature scale. If we are interested in the piece of de Sitter associated to a single observer, or more precisely the intersection of all future events and past events of an inertial worldline, we find a rather different description of the spacetime, namely
\begin{equation}\label{static}
\frac{ds^2}{\ell^2} = -dt^2 \cos^2 \rho + d\rho^2 + \sin^2\rho \, d\varphi^2~,
\end{equation}
with $\varphi\sim \varphi+2\pi$, $t\in\mathbb{R}$, and $0 \leq \rho \leq \pi/2$. The piece of global de Sitter that this static metric covers is the shaded region in figure \ref{fig:SoftdeSitterPenrose}. The exponentially expanding character of (\ref{global}) is subsumed into the presence of a cosmological event horizon of size $2\pi \ell$ residing at $\rho=\pi/2$. An inertial observer is described by the time-like curve $\rho=0$, and the light emitted toward them by objects approaching $\rho = \pi/2$ becomes increasingly redshifted. Finally, we note that the original global Cauchy surface at $\mathcal{T}=0$, which takes the form of a spatial $S^2$, has been split two disks whose common $S^1$ boundary comprises the dS$_3$ horizon. The induced metric on each disk at a constant $t$ slice is given by
\begin{equation}\label{diskmetric}
\frac{ds^2}{\ell^2} = d\rho^2 + \sin^2\rho \, d\varphi^2~.
\end{equation}
It is occasionally convenient to view dS$_3$ as the metric induced  on a hypersurface embedded in $\mathbb{R}^{1,3}$. Denoting the coordinates of $\mathbb{R}^{1,3}$ by $(X^0,\bold{X})$, the hypersurface is described by the following equation
\begin{equation}\label{embedding}
-(X^0)^2 + \bold{X} \cdot \bold{X} = \ell^2~.
\end{equation}
The hypersurface preserves an $SO(3,1) \cong SL(2,\mathbb{C})/\mathbb{Z}_2$ subgroup of the Poincar{\'e} symmetries of $\mathbb{R}^{1,3}$. This subgroup constitutes the isometry group of dS$_3$. We can recover the metrics (\ref{global}) and (\ref{static}) through specific parametrisations for the $(X^0,\bold{X})$. 

Another geometric feature clear from (\ref{embedding}) is the relation of dS$_3$ to the round metric on $S^3$. Indeed, upon analytically continuing $X^0 \to -i X^0$ the equation (\ref{embedding}) becomes that of an $S^3$ embedded in $\mathbb{R}^4$, and the isometries become $SO(4)\cong SU(2)\times SU(2)/\mathbb{Z}_2$. From the perspective of our two charts, the continuation corresponds to $\mathcal{T} \to - i \theta$ in (\ref{global}), and $t \to -i \xi$ in (\ref{static}). If the latter continuation is to give a smooth Euclidean space, we must further periodically identify $\xi \sim \xi + 2\pi$. The periodicity of the Euclidean static patch time coordinate $\xi$ lies at the heart of the thermal properties of the dS$_3$ horizon. From it we can read off the temperature of the dS$_3$ horizon, $T_{\text{dS}} = ( 2\pi\ell )^{-1}$, as measured by an inertial observer at $\rho=0$. It is somewhat remarkable that both the global and static patches, which describe rather distinct pieces of the Lorentzian space-time, analytically continue to the same compact space in Euclidean signature. The `observer independence' of the de Sitter horizon is reflected in Euclidean signature by the fact that any great circle on the $S^3$ can be viewed as the Euclidean de Sitter horizon.

Given the presence of a horizon, it is natural to consider quantum fields in a thermal density matrix within a single static patch. This can be achieved by placing them in a pure Hartle-Hawking state across the full $S^2$ spatial slice at $\mathcal{T}=0$ and tracing out one of the two static patch regions. In much of what follows we will consider contributions to the thermal properties of the static patch horizon from Chern-Simons and gravitational fields. Before doing so we end a few more general comments.  

\subsection{Entanglement $\&$ gravitational entropy}\label{graventropy}

It is generally a complicated problem to separate two regions of a gauge theory (including gravitational theories) in a strictly local way. In discussions of entanglement entropy for ordinary gauge theories there are various proposals for how to deal with the problem of the entangling surface \cite{Buividovich:2008gq,Donnelly:2011hn,Donnelly:2014gva,Casini:2013rba,Ghosh:2015iwa,Lin:2018bud}. For standard Yang-Mills theories, the choices are often labelled by a choice of `centre'  of the gauge invariant operator algebra. In Chern-Simons theories, which we will proceed to study, electric and magnetic fields do not commute. Consequently, it is not reasonable to classify choices in terms of electric and magnetic centres. Alternatively, \cite{Buividovich:2008gq,Donnelly:2014gva} one might extend the gauge invariant Hilbert space to admit non-trivial charges at the entangling surface. As overviewed in \cite{Lin:2018bud}, the extended Hilbert space picture matches calculations of entanglement entropy employing Euclidean path integral techniques. Our considerations will be in line with this approach, since we are interested in making contact with the manifestly gauge-invariant three-sphere partition function. 

Consider Chern-Simons theory with gauge group $\mathcal{G}$ and the level $k$ and whose three-sphere partition function we denote by $Z_{\mathcal{G}_k}[S^3]$.  Although $Z_{\mathcal{G}_k}[S^3]$ is a manifestly gauge invariant object it will be related to a Lorentzian edge-mode calculation, which we describe in sections \ref{css3} and \ref{csab}, for which gauge invariance must be manifestly broken. Both the entanglement entropy of the edge-mode theory as well as $Z_{\mathcal{G}_k}[S^3]$ are ultraviolet divergent quantities. We will make sense \cite{Anninos:2020hfj} of the divergences of $Z_{\mathcal{G}_k}[S^3]$ by coupling the theory to three-dimensional gravity with cosmological constant $\Lambda = 1/\ell^2>0$. The low energy effective theory in Euclidean signature is thus given by\footnote{The Chern-Simons action (being insensitive to the metric) is insensitive to the signature and hence appears in the same form for both the Lorentzian and Euclidean path integral.}
\begin{equation}
-S_E[g_{\mu\nu},A_\mu] = \frac{1}{16\pi G} \int d^3x \sqrt{g} \left( R-\frac{2}{\ell^2} \right) + i S_{\text{CS}}[A_\mu]~,
\end{equation}
where $S_{\text{CS}}[A_\mu]$ is the Chern-Simons action. Unlike pure Chern-Simons theory, the above theory is no longer topological due to the explicit metric dependence in the gravitational sector. The classical solution space for the metric remains that of three-dimensional gravity due to the metric independence of Chern-Simons theory. In three-dimensions, this is nothing more than the round metric on the three-sphere or quotients thereof. The path integral of interest now becomes
\begin{equation}
\mathcal{Z}_{\text{grav} + \text{CS}} = Z^{\text{reg}}_{\mathcal{G}_k}[S^3] \times \mathcal{Z}_{\text{grav}}~.
\end{equation}
The pure gravitational path integral $\mathcal{Z}_{\text{grav}}$ is given by (\ref{logZgrav}). Any ultraviolet divergences stemming from quantum fluctuations of the Chern-Simons sector can be absorbed into the  cosmological constant $\Lambda$ and Newton constant $G$, such that $Z^{\text{reg}}_{\mathcal{G}_k}[S^3]$ is the finite ultraviolet regularised part of the Chern-Simons partition function. The overall phase of $Z^{\text{reg}}_{\mathcal{G}_k}[S^3]$ is affected by the choice of framing and is given by $\varphi_{\text{CS}} = 2 \pi n c_k/24$ where $c_k$ is the left-moving central charge of the WZW model with current algebra $\mathcal{G}_k$ \cite{witten1989quantum} --- for $SU(2)_k$ this would be $c_k =3k/(k+2)$ --- and $n\in\mathbb{Z}$. We note that the total phase of $\mathcal{Z}_{\text{grav} + \text{CS}}$ is then given by
\begin{equation}
\varphi_{\text{grav} + \text{CS}} = \varphi_{\text{grav}} + \frac{2 \pi n c_k }{24} + \ldots 
\end{equation}
For the gravitational phase $\varphi_{\text{grav}} = -5\pi/2$ proposed in \cite{Polchinski:1988ua}, special choices of $n$ and $k$ can lead to a vanishing $\varphi_{\text{grav} + \text{CS}}$ modulo $2\pi$, at least to leading order in the large $S_{\text{dS}}$ expansion. 

At least for vanishing phase, $\log Z^{\text{reg}}_{\mathcal{G}_k}[S^3]$ can be viewed as a physical, ultraviolet finite correction to the tree-level Gibbons-Hawking de Sitter horizon entropy $S_{\text{dS}} = \pi\ell/2G$ \cite{Gibbons:1976ue}. So long as it is small compared to $S_{\text{dS}}$, there is no issue with the contribution to the entanglement entropy from the Chern-Simons sector being negative.

\section{Chern-Simons theory on dS$_3$ and $S^3$}\label{css3}

In this section we consider Chern-Simons theory on the manifolds $S^3$, $S^2\times \mathbb{R}$, and $D\times \mathbb{R}$, where $D$ denotes the two-dimensional disk. We frame the discussion in the language of Lorentzian and Euclidean three-dimensional de Sitter space. Chern-Simons theory with gauge group $\mathcal{G}$ and level $k \in \mathbb{Z}^+$ is described by the action
\begin{equation}\label{CSaction}
S_{\text{CS}}[A_\mu]  
= \frac{k}{4 \pi}   \int_M d^3 x \, \varepsilon^{\mu \nu \rho} \, \text{Tr} \,\left(A_{\mu}\partial_{\nu}A_{\rho}  + \frac{2}{3}A_{\mu}A_{\nu} A_{\rho}\right)~,
\end{equation}
with $A_\mu = A_{\mu}^a T^a$, where $T^a$ are the anti-Hermitian generators of $\mathcal{G}$ satisfying the normalisation $\text{Tr}(T^a T^b) = - \frac{1}{2} \delta^{ab}$, and $a = 1,..., \text{dim}\,\mathcal{G}$, while $ \varepsilon^{\mu \nu \rho}$ is the Levi-Civita symbol. The non-Abelian field strength tensor is $F_{\mu\nu} = \partial_{\mu}A_{\nu} - \partial_{\nu}A_{\mu} + [A_{\mu},A_{\nu}]$. 

Our reason for studying Chern-Simons theories is threefold. Firstly, the topological nature of Chern-Simons theory ensures that all excitations have vanishing energy -- all physics is `soft physics'. In particular, it contributes to the non-local structure of the de Sitter entropy (\ref{logZgrav}). Secondly, we can draw from a host of exact results for Chern-Simons theory on compact manifolds, as well as on manifolds with boundaries. Both these features will prove to be particularly relevant to questions regarding de Sitter space. Thirdly, general relativity in three-dimensions is equivalent, at least at the semiclassical level, to a Chern-Simons theory exhibiting certain unusual features such as a non-compact gauge group and/or a complexified value for the level \cite{Achucarro:1987vz,witten19882+}. For now, we will steer clear of the third point, which we return to in section \ref{oneloop3dgrav}, and focus on the first two.

\subsection{Chern-Simons theory on $S^3$}

We now consider the behaviour of Chern-Simons gauge fields on a fixed dS$_3$ or $S^3$ background. Pure Chern-Simons theory on a three-manifold $\mathcal{M}$ will only perceive topological features, so one might wonder if there is any lesson to be learned specifically about de Sitter. Here,  as  was discussed  in section \ref{general}, it is helpful to consider the more general setup whereby we couple Chern-Simons theory to three-dimensional gravity. The Lorentzian/Euclidean theory admits dS$_3$/$S^3$ solutions. Given that the metric is semiclassically fixed, the theory is no longer topological. In this context we can ask how the presence of a Chern-Simons gauge field contributes to the thermodynamic properties of the dS$_3$ horizon. 

In the remainder of this section and the next our focus will be entirely on the case of a fixed background. We take the gauge group $\mathcal{G}$ to be compact and semi-simple, and the level $k \in \mathbb{Z}^+$. Recall that Chern-Simons theory is topological, and hence only sensitive to the type of manifold on which it resides. From our discussion of the dS$_3$ geometry one is led to consider various three manifolds. These are $S^3$, $S^2\times \mathbb{R}$, and $D \times \mathbb{R}$. 
\newline\newline
\textbf{Chern-Simons theory on $S^3$.} The first object we consider is the partition function of Chern-Simons theory on $S^3$. Although na{\"i}vely this is nothing more than a number, and hence as good as our choice of normalisation, it turns out that Chern-Simons theory is sufficiently structured that one can make sense of this number in an essentially unambiguous way.\footnote{There is an ambiguity in the overall phase of the $S^3$ partition function, which is fixed by a choice of framing. Unless otherwise stated, we will select a framing for which the overall phase vanishes.} The result  \cite{witten1989quantum} can be stated concisely in terms of the modular $S$-matrix ${\mathcal{S}_m}^n$ associated to the WZW CFT with gauge group $\mathcal{G}$ and level $k$, namely
\begin{equation}
Z_{\mathcal{G}_k}[S^3] = {\mathcal{S}_0}^0~.
\end{equation}  
(An account of modular $S$-matrices and their properties can be found in the latter chapters of \cite{difrancesco}. They encode the transformation of extended characters under inversion.) It should be emphasised that $Z_{\mathcal{G}_k}[S^3]$ is a gauge invariant object. As a concrete example, we can consider $\mathcal{G} = SU(5)$ for which
\begin{equation}
Z_{SU(5)_k}[S^3] = \frac{1}{\sqrt{5}(k+5)^2} \prod_{j=1}^{4} \left[2\sin \frac{\pi j}{k+5}\right]^{(5-j)}~.
\end{equation}  
Notice that in the perturbative limit, where $k \to \infty$, the above expression is approximated by 
\begin{equation}
\lim_{k\to\infty} Z_{SU(5)_k}[S^3] \approx \frac{9}{\pi ^{14} \, 512\sqrt{5}}   \left(\frac{2\pi}{\sqrt{k+5}}\right)^{24}~.
\end{equation}
The above expression has some identifiable features. For instance the power of $\sqrt{k+5}$ corresponds to $\text{dim} \,SU(5)=24$. Whereas the numerical pre-factor is the reciprocal of the canonical group volume of $SU(5)$, given by (see for example \cite{ooguri2002worldsheet,anninos2020notes}) 
\begin{equation}
\text{vol} \, SU(N) = \frac{\sqrt{N}}{2\pi} \frac{(2\pi)^{N(N+1)/2}}{G(N+1)}~,
\end{equation}
where $G(z)$ is the Barnes $G$-function. 

More generally, the large $k$ limit of the sphere partition function of Chern-Simons theory with gauge group $\mathcal{G}$ is given by 
\begin{equation}\label{CSpert}
\lim_{k\to\infty} Z_{\mathcal{G}_k}[S^3] \approx \frac{1}{\text{vol} \, \mathcal{G}} \left( \frac{2\pi}{\sqrt{k+h}}\right)^{\text{dim}\, \mathcal{G}}~,
\end{equation} 
where $h$ is the dual Coxeter number of $\mathcal{G}$. Schematically, we may view the above structure as stemming from the fact that the path integral requires division by the space of gauge transformations, of which all but the constant part are cancelled by redundancies in the local description of the theory. When a quantum field theory lives on a compact space, the constant part of a compact gauge group is finite and must be taken into account. Indeed, its contribution is responsible for the group theoretic factors exhibited by (\ref{CSpert}). The dependence on the level $k$ is also sensible. The volume of the constant part of the gauge group must be normalised with respect to the size of the fluctuations of the theory, which are controlled by $1/\sqrt{k}$ in the perturbative limit. Such group theoretical factors have been recently explored in \cite{Anninos:2020hfj}. 

\subsection{Chern-Simons on $S^2\times \mathbb{R}$ and $D \times \mathbb{R}$}

\begin{figure}[H]
	\centering
	\includegraphics[width=0.37\linewidth]{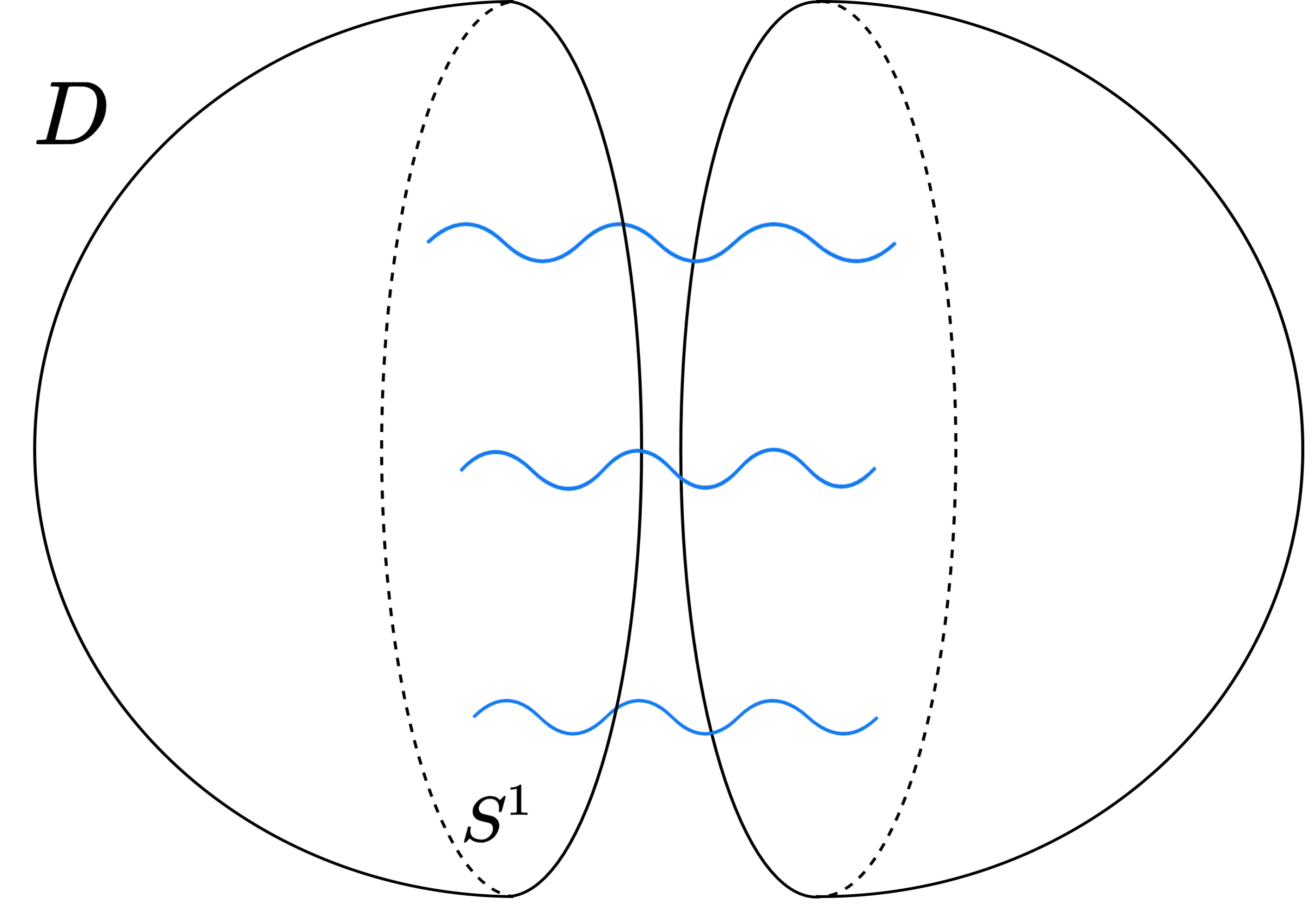}
	\caption{The $\mathcal{T}=0$ slice of global de Sitter split into two static patch hemispheres at the $S^1$ horizon. These hemispheres are topologically disks, and we can imagine `tracing out' the degrees of freedom in one of the static patches that is behind the horizon of the other.}
	\label{fig:S2tracingoutEE}
\end{figure}

We now consider Chern-Simons theory on $S^2\times \mathbb{R}$ and $D \times \mathbb{R}$. The first is topologically equivalent to the global geometry (\ref{global}) of dS$_3$, while the second is topologically equivalent to the static patch (\ref{static}). 
\newline\newline
\textbf{{Chern-Simons theory on $S^2\times \mathbb{R}$}.} We now proceed to consider Chern-Simons theory on $S^2\times \mathbb{R}$. Here, we view $\mathbb{R}$ as the global dS$_3$ time coordinate $\mathcal{T}$, and the $S^2$ as the spatial Cauchy slice of global dS$_3$. It is convenient to choose the following metric on $S^2$
\begin{equation}
ds^2 = \frac{4 d \bold{x}^2}{(1+\bold{x}^2)^2}~, \quad\quad \bold{x} = \{x,y\} \in \mathbb{R}^2~.
\end{equation}
We begin with a description of the Hilbert space. This is given by quantising the space of flat $\mathcal{G}$-connections on $S^2$, modulo gauge transformations. But on $S^2\times \mathbb{R}$, all flat connections are trivial. The Hilbert space is one-dimensional. The unique state of this peculiar world can be explicitly described in the Schr{\"o}dinger picture. To describe it, it is convenient to work in the gauge $A_\mathcal{T}=0$. The gauge constraints that we must solve are then given by
\begin{equation}
F_{ij} = 0~, \quad\quad i,j \in \{x,y\}~.
\end{equation}
The following quantum state can be shown to solve the above equation \cite{dunne1989chern}
\begin{equation} \label{CSnon-abelianWavefucntion}
\Psi_{S^2}[A_x(\bold{x})] = \mathcal{N} \exp{\left[\frac{ik}{4\pi}  \text{Tr} \int_{S^2} d^2 {x}   \left[ (g^{-1}\partial_x g) ( g^{-1} \p_y g) \right] - 2 \pi i k \int_{S^2} \, d^2 {x} \, w^0 (g) \right]}~,
\end{equation}
where $\mathcal{N}$ is a normalisation constant, $g(\bold{x}) \in \mathcal{G}$, and $w^0$ is the time-component of the three-vector whose divergence gives the winding number density:
\begin{equation}
W(g) = \partial_{\mu} w^{\mu} = \tfrac{1}{24 \pi^2} \text{Tr} \, \varepsilon^{\alpha \beta \gamma} (g^{-1} \partial_{\alpha}g g^{-1} \partial_{\beta}g g^{-1} \partial_{\gamma}g)~.
\end{equation}
The gauge field is non-locally related to $g(\bold{x})$ as $A_x(\bold{x}) = g^{-1} \partial_x g$. 
Unlike what happens in standard Yang-Mills theory, the Chern-Simons wavefunctional $\Psi_{S^2}[A_x(\bold{x})]$ is not gauge invariant under the residual gauge freedom 
\begin{equation}
A^h_i(\bold{x}) = h^{-1}\left( \partial_i + A_i(\bold{x})\right) h~, \quad\quad i \in \{x,y\}~,
\end{equation}
with $h(\bold{x}) \in \mathcal{G}$. Rather, under the residual gauge freedom
$\Psi_{S^2}[A_x(\bold{x})]$ transforms as 
\begin{equation}
\Psi_{S^2}[A_x(\bold{x})] = e^{2\pi i \alpha(A_x(\bold{x}); \,h)} \Psi_{S^2}[A_x(\bold{x})]~,
\end{equation} 
where
\begin{equation}
\alpha (A_x(\bold{x}); h) \equiv \frac{k}{8\pi^2}  \text{Tr}  \int d^2 x \left[2A_x(\bold{x}) \, \partial_y h \, h^{-1} + h^{-1}\partial_x h \, h^{-1} \, \partial_y h \right] - k \int d^2 x \, w^0 (h). 
\end{equation}
As a function of the gauge field the ground state $\Psi_{S^2}[A_x(\bold{x})]$ encodes non-local quantum correlations. The non-local correlations play an important role, for instance, if we are to trace out those degrees of freedom living in a certain spatial region. Our considerations of dS$_3$ suggest performing precisely such a trace if we are to consider the physics contained within a single dS$_3$ horizon, as depicted in Figure \ref{fig:S2tracingoutEE}. 
\noindent {\textbf{Chern-Simons theory on $D \times \mathbb{R}$.}} As already noted, the spatial section of a single static patch is given by the hemisphere metric on a two-dimensional disk $D$ whose coordinates we denote by $\{\rho,\varphi\}$ with $\varphi\sim\varphi+2\pi$. From this perspective, we are encouraged to consider the problem of Chern-Simons theory on $D\times \mathbb{R}$, where now $\mathbb{R}$ is associated with the static patch time $t$. In this case, the Hilbert space of Chern-Simons theory is far richer \cite{Elitzur:1989nr}  due to the presence of the spatial $S^1$ boundary. 

We can construct the edge-mode theory as follows (see \cite{tong} for an overview). First, we must ensure that we have a well-defined variational problem in the presence of the $S^1$ boundary. We will not add further terms localised on the boundary $\partial\mathcal{M}$ of the disk. As such, the variational problem enforces us to impose boundary conditions on $\partial \mathcal{M}$:
\begin{equation}\label{gauge}
\left( A_t - \upsilon A_\varphi \right)|_{\partial \mathcal{M}} = 0~.
\end{equation}
The parameter $\upsilon$ is a real number which we take to be positive. We must further ensure that the gauge transformations do not disturb our boundary condition. This can be achieved by forcing the gauge parameter $\lambda$ to vanish on the $S^1$ boundary. We can further impose (at least locally) that the gauge condition (\ref{gauge}) holds away from the boundary of the disk as well. For that to be the case the bulk gauge parameter must  satisfy
\begin{equation}
\left(\partial_t -\upsilon \partial_\varphi \right) \lambda =  0~, \quad\quad \lambda|_{S^1} = 0~.
\end{equation} 
The above is solved by $\lambda = \bar{\lambda}(\varphi + \upsilon t,\rho)$ with $\bar{\lambda}|_{S^1} = 0$, which constitutes a residual gauge freedom. We can further specify a gauge fixing condition throughout the interior of the disk. The constraint arising from extending the gauge condition (\ref{gauge}) into the disk interior is solved by the following configurations
\begin{equation}
A_\rho  =  U^{-1} \partial_\rho U~, \quad\quad   A_\varphi  =  U^{-1} \partial_\varphi U~,
\end{equation}
with $U \in \mathcal{G}$. The residual gauge freedom transforms the group valued function $U \to e^{i \varepsilon \bar{\lambda}} U$, where $\varepsilon$ is a small parameter. Consequently, the residual gauge parameter can be entirely fixed upon fixing the form of $U$. Unless otherwise specified, we impose that $A_\mu$ remains smooth throughout the interior of the disk. To do so, we must ensure that the holonomy of the gauge field vanishes around any closed loop in the interior of the disk. Non-vanishing holonomy would indicate the presence of external charge puncturing the disk. 
The resulting theory at the boundary of the disk, upon taking everything into account, becomes a chiral WZW model with group $\mathcal{G}$ and level $k$ built from the boundary values of $U$ \cite{Elitzur:1989nr}. Upon quantisation, this theory has an infinite number of states. We now study the Abelian case in detail.

\section{Abelian example}\label{csab}

In this section we construct and explore the edge-mode theory associated to Abelian Chern-Simons theory with gauge group $U(1)$ at level $k$ quantised on a spatial disk $D$. The purpose of the section is to provide a detailed discussion of this edge-mode theory and its relation to the three-sphere partition function $Z_{U(1)_k}[S^3]$. This is meant to serve as the basic template for the discussion of edge-mode theories in the latter sections. 

The Abelian theory on a three-manifold $\mathcal{M}$ is governed by the action
\begin{equation}
S_{U(1)_k}[A_\mu] = \frac{k}{4\pi} \int_{\mathcal{M}} d^3 x \, \varepsilon^{\mu \nu \rho} A_\mu \partial_\nu A_\rho~,
\end{equation}
where $A_\mu$ is a real-valued Abelian gauge field. So long as the three-manifold $\mathcal{M}$ has no boundary, the above is gauge invariant under $A_\mu \to A_\mu - \partial_\mu \lambda$, where $\lambda$ is a smooth function on $\mathcal{M}$. If it is further assumed that $U(1)$ is compact, and moreover that $S_{U(1)_k}$ is the low energy limit of a theory containing both electric and magnetic monopoles, then the standard Dirac argument enforces $k\in\mathbb{Z}$. For the sake of simplicity, we  take $k \in 2  \times \mathbb{Z}^+$ in what follows.\footnote{This is, in part, to avoid subtleties associated to fermionic states in the edge-mode theory.}

\subsection{Three-sphere partition function} 

We begin by computing the $S^3$ partition function. We have 
\begin{equation}
Z_{U(1)_k}[S^3] =  \left( {\text{vol} \, \mathcal{G}}\right)^{-1} \, {\int \mathcal{D}A_\mu \, e^{i S_{U(1)_k}[A_\mu]}}~,
\end{equation}
where we have manifestly divided by the volume of the gauge group, which is generated by the space of smooth real functions on $S^3$. Given that $S^3$ is compact, we must also consider the constant part of the gauge group which is generated by the constant function $\lambda_c$ on $S^3$. We take the group elements generated by $\lambda_c$ to be given by
\begin{equation}
U(\lambda_c) = e^{i \lambda_c}~,
\end{equation}
such that $\lambda_c$ is compact with radius $2\pi$. It is convenient to rescale the gauge field by $\sqrt{2\pi/k}$, to remove the $k$ dependence from the action. In doing so, we change the volume of the constant part of the gauge group to $\sqrt{2\pi k}$. Whenever we evaluate path integrals, we will assume this normalisation for the action. To get a feel for the structure of the result \cite{Giombi:2015haa}, we can parameterise $A_\mu$ as
\begin{equation}
A_\mu = {A}^T_\mu + \partial_\mu \mathcal{B}~, \quad\quad \nabla^\mu {A}^T_\mu= 0~,
\end{equation} 
where we exclude the constant part of the scalar field $\mathcal{B}$ as it does not contribute to the configuration space of the $A_\mu$. We thus find
\begin{equation}\label{unfixed}
Z_{U(1)_k}[S^3] =  \left( {\text{vol} \, \mathcal{G}}\right)^{-1} \times \sqrt{{\det}' ( -\nabla^2)} \times \int \mathcal{D}'\mathcal{B}  \int \mathcal{D}{A}^T_\mu \, e^{i S_{U(1)_k}[{A}^T_\mu]}~.
\end{equation}
The prime indicates we are dropping the zero-mode. Although the above expression is still rather schematic, one can already note that the path-integral over $\mathcal{B}$ will mostly cancel the volume of $\mathcal{G}$. The difference lies in the zero-mode sector which is absent in the space of functions $\mathcal{B}$. Since no other term in the partition function depends on $k$, we can already conclude that
\begin{equation}
Z_{U(1)_k}[S^3]  \propto \sqrt{\frac{1}{{k}}}~.
\end{equation}
We now fix the remaining proportionality constant. One way to do so involves calculating and regularising the divergent functional determinants stemming from the Gaussian path integral. Generally speaking, a local quantum field theory in three-dimensions on a compact space with metric $g_{\mu\nu}$ will have a partition function of the form
\begin{equation}\label{divergence}
\log Z[g_{\mu\nu}] =  c_0 \, \ell_{\text{uv}}^{-3} \int d^3x \sqrt{g} + c_1 \, \ell_{\text{uv}}^{-1} \int d^3x \sqrt{g} R + \text{finite}~,
\end{equation}
where $\ell_{\text{uv}}$ is an ultraviolet length scale. The finite piece will encode some non-local functional of $g_{\mu\nu}$.  If the theory is not parity invariant, as is the case for Chern-Simons theory, one will generally have a local contribution proportional to the gravitational Chern-Simons term contributing to the $\Lambda$-independent part. In certain circumstances, due to the presence of additional symmetries or structures, such as supersymmetry, some of the divergences can be argued to be absent. As we discuss in appendix \ref{S3det}, applying the Fadeev-Popov procedure in the Lorenz gauge to the Abelian Chern-Simons theory yields the following contribution to the sphere partition function
\begin{equation}\label{Zdet}
Z_{U(1)_k}[S_3] = e^{i \varphi_{\text{CS}}} \left({\text{vol} \,S^3\, \ell_{\text{uv}}^{-3}}\right)^{-1/2}  \, \sqrt{\frac{1}{k}} \times \frac{\left({\det' \left(- \nabla^2 /\ell_{\text{uv}}^{-2}\right)}\right)^{1/2}}{\left(\det L L^\dag /\ell_{\text{uv}}^{-2} \right)^{1/4}}~,
\end{equation}
where $L$ is the operator $i\varepsilon^{\mu\nu\rho}\partial_\rho$ acting on the space of transverse vector fields, and the numerator stems from path integration over the ghost fields. The prime in the determinant means we are omitting any zero modes, which must be dealt with separately. Any of the local ultraviolet divergences in (\ref{divergence}) come from evaluating the functional determinants in (\ref{Zdet}). 

To assess the nature to the leading cubic divergence, it is sufficient to consider the problem on $\mathbb{R}^3$ (or a large enough box). In this case the spectra are straightforward to obtain, and one finds that the eigenvalues of both $L L^\dag$ and $-\nabla^2$ are given by $\bold{k}\cdot\bold{k}$ with $L L^\dag$ having twice the multiplicity due to the two polarisations of $A^T_\mu$. Consequently, at least the absolute value of $Z_{U(1)_k}[S_3]$ is free of cubic ultraviolet divergences. This resonates well with the absence of local degrees of freedom in Chern-Simons theory. So far, our arguments do not suffice to conclude anything about the linear divergence of $Z_{U(1)_k}[S_3]$.  As we discuss in appendix \ref{S3det}, a linear divergence is indeed present in the heat kernel regularisation scheme. We find
\begin{equation}\label{AbelianZ}
|Z_{U(1)_k}[S^3]| = \sqrt{\frac{1}{k}} e^{-\frac{3\pi}{4\varepsilon}}~, \quad\quad \varepsilon \equiv \frac{2 e^{-\gamma}\ell_{\text{uv}}}{\ell}~.
\end{equation}
The phase $\varphi_{\text{CS}}$ of $Z_{U(1)_k}[S_3]$ requires a careful treatment. We can heuristically argue that it will not contribute to the cubic divergence either. Indeed, given that the phase is associated to the parity non-invariance of the theory, we might expect any associated divergence to also be parity non-invariant such as the gravitational Chern-Simons term.\footnote{Alternatively, we could consider adding two Chern-Simons terms with equal and opposite level, and regularise in such a way that the phase of the sphere partition function vanishes.}

A more sophisticated approach to compute the sphere partition function, and its phase, follows \cite{witten1989quantum}. Since this approach will be of use more generally, we now discuss it. Consider two solid tori, $\bold{T}_L^2 = D_L\times S^1$ and $\bold{T}_R^2 = D_R\times S^1$, with boundary tori $T^2_L$ and $T^2_R$. If we glue these together by identifying the points on $T_L^2$ with those on an oppositely oriented $T_R^2$,  we obtain an $S^2 \times S^1$. The partition function $Z_{\mathcal{G}_k}[S^2\times S^1]$ counts the number of states of Chern-Simons theory with gauge group $\mathcal{G}$ at level $k$ quantised on a spatial $S^2$, that is to say
\begin{equation}
Z_{\mathcal{G}_k}[S^2\times S^1] \equiv 1~.
\end{equation} 
That the above equation holds, regardless of the gauge group and level (so long as there are no punctures on the $S^2$) is simply the statement that Chen-Simons theory on a spatial $S^2$ has a unique state in its Hilbert space. What is perhaps less immediate is that by identifying the points on $T^2_L$ with a particular modular transformation $\tau_R$ of the points on $T^2_R$, we instead get an $S^3$. This is true for a variety of $\tau_R$. For our immediate purpose we will consider an inversion $\tilde{\tau}_R  = -1/\tau_R$. Finally, we must now take into account the general relation 
\begin{equation}\label{ZSmod}
Z_{\mathcal{G}_k} [\tilde{\bold{T}}^2] = {\mathcal{S}_0}^0 \times Z_{\mathcal{G}_k} [\bold{T}^2 ]~,
\end{equation}
where $\tilde{\bold{T}}^2$ is a solid torus whose boundary is given by that of $\bold{T}^2$ upon performing the inversion $\tilde{\tau}  = -1/\tau$. The above formula follows from a careful consideration of the solid torus Hilbert space \cite{witten1989quantum}, but we will not attempt to derive it here. If we can view the partition functions on the left and right hand sides of (\ref{ZSmod}) as quantum wavefunctions in the Hilbert space of the theory quantised on a spatial $T^2$, we can take an inner product of both sides with the state on the right hand side to get
\begin{equation}
{Z_{U(1)_k} [S^3]}  = {\mathcal{S}_0}^0 \times Z_{U(1)_k}[S^2 \times S^1]~.
\end{equation}
Using the modular $S$-matrix for $U(1)_k$, it is concluded that
\begin{equation}\label{Zu1}
{Z_{U(1)_k} [S^3]} = \sqrt{\frac{1}{k}}~.
\end{equation}
In such a way, we can fix the overall constant and phase in (\ref{unfixed}). We now proceed to consider the above expressions from the perspective of the theory quantised on a spatial disk. 

\subsection{Classical edge-mode theory}\label{edgemodeu1}

Quantisation of Chern-Simons theory on a spatial disk requires a careful consideration of boundary conditions on the $S^1$ boundary. As already mentioned, a well-posed variational problem with Dirichlet conditions on the gauge field enforces the boundary condition (\ref{gauge}). Imposing the condition (\ref{gauge}) throughout the remainder of space fixes our gauge entirely. The resulting constraints impose that the spatial components of the gauge field take the following form
\begin{equation}\label{constrain}
A_\rho  = i \, e^{i \, \Xi(t,\varphi,\rho)} \, \partial_\rho \, e^{-i \, \Xi(t,\varphi,\rho)}~, \quad\quad {{A_\varphi   = i \, e^{i \, \Xi(t,\varphi,\rho)} \partial_\varphi e^{-i \, \Xi(t,\varphi,\rho)}}}~,
\end{equation}
where we recall that $\rho\in[0,\pi/2]$ and $\varphi \sim \varphi + 2\pi$ are coordinates on the disk (\ref{diskmetric}). 

The edge-mode theory residing at the boundary of the disk where $\rho=\pi/2$, in the case of $\mathcal{G} = U(1)$, is described by the Floreanini-Jackiw action \cite{Floreanini:1987as}
\begin{equation}\label{edge}
S_{\text{edge}} = \frac{k}{4\pi} \int d t d\varphi \, \partial_\varphi \zeta  \left( \partial_t  - \upsilon \partial_\varphi  \right) \zeta~,
\end{equation}
where the field $\zeta(t,\varphi) = \Xi(t,\varphi,\rho)|_{\rho=\pi/2}$ is a compact scalar with periodicity $\zeta \sim \zeta + 2\pi$ field mapping $\mathbb{R} \times S^1 \to S^1$. The above theory is that of a compact chiral boson. Classically, the solution space is given by
\begin{equation}\label{solns}
\zeta(t,\varphi) = g(\varphi + \upsilon t) +  m\, \varphi + f(t)~.
\end{equation}
The first term corresponds to a chiral excitation moving at angular velocity $\upsilon$ \cite{Wen:1990se}. The second term, which arises due to the compactness of $\zeta$, provides a winding number $m\in\mathbb{Z}$ counting the number of times $\zeta$ wraps around the spatial $S^1$. 
The third term does not contribute to the classical energy of the solution, as can be seen from the classical Hamiltonian
\begin{equation}
H = \frac{k\upsilon}{4\pi}\int d\varphi \, (\partial_{\varphi}\zeta)^2~.
\end{equation}
As can be seen from (\ref{constrain}), $f(t)$ carries no physical information and can be dropped all together. The global $U(1)$ symmetry of the edge-mode theory corresponds to shifts of $\zeta$ (modulo integer multiples of $2\pi$), and is generated by the following charge
\begin{equation}\label{U1charge}
\mathcal{Q} = \int \frac{d\varphi}{2\pi} \partial_\varphi \zeta~.
\end{equation}
It is convenient to expand the non-winding mode sector in a Fourier expansion
\begin{equation}
\zeta(t,\varphi) = \frac{1}{\sqrt{2\pi}} \sum_{n \in\mathbb{Z}^+} \left(\alpha_n(t) e^{-i n \varphi} + \bar{\alpha}_n(t) e^{i n \varphi} \right)~.
\end{equation}

\subsection{Quantum edge-mode theory}

Upon quantisation, the complex functions $\alpha_n(t)$ are promoted to operators satisfying the equal-time commutation relations
\begin{equation}\label{commutators}
[\hat{\alpha}_n,\hat{\alpha}_m]  = [\hat{\alpha}_n^\dag,\hat{\alpha}_m^\dag]  =0~, \quad\quad\quad [\hat{\alpha}_n,\hat{\alpha}_m^\dag] =  \frac{2\pi}{k n}  \, \delta_{m,n}~.
\end{equation}
To derive the above, we should keep in mind that this problem involves a constrained phase space. It follows from the commutation relations (\ref{commutators})  that the edge-mode theory contains a level $k$ $\mathfrak{u}(1)$ Kac-Moody algebra. The periodicity $\zeta \sim \zeta  + 2\pi$ leads us to consider the vertex operators
\begin{equation}\label{Psi}
\hat{\mathcal{O}}_n = \, : e^{i n \hat{\zeta}} :~, \quad\quad n = 1,2,\ldots,k-1~.
\end{equation}
Acting with the $\hat{\mathcal{O}}_n$ inserts a timelike Wilson line piercing the interior of the disk. The operator carries fractional charge $\mathcal{Q}_n = n/k$ under the $U(1)$ shift symmetry (\ref{U1charge}) of the edge-mode theory. We view such anyonic insertions as external data to the edge-mode theory, and consequently not part of the edge-mode Hilbert space. For $n=m k$, with $m\in\mathbb{Z}$, the above operators are no longer singular in the disk interior. Indeed, the closed-loop integral $\oint A_\varphi d\varphi$ from a single winding mode evaluates to $2\pi m$. Instead, for  $n=m k$ these correspond to winding mode excitations which indeed reside in the edge-mode Hilbert space.

The quantum Hamiltonian for a given winding sector $m$ is given by 
\begin{equation}\label{eHam}
\hat{H}_m = k \, \upsilon \frac{m^2}{2} \hat{\mathbb{I}}_m  +\frac{\upsilon k}{2\pi}  \sum_{n \in\mathbb{Z}^+} n^2 \, \hat{\alpha}^\dag_n \hat{\alpha}_n~,
\end{equation}
where $\hat{\mathbb{I}}_m$ is the identify operator in a  given winding sector. 
The ground state is given by the state carrying vanishing winding and annihilated by all the $\hat{\alpha}_n$. In a given winding sector the eigenstates of the Hamiltonian are given by
\begin{equation}
|n_1,d_1;n_2,d_2;\ldots,n_p,d_p \rangle_m = \prod_{i=1}^p  \sqrt{\frac{1}{d_i !} \, \left(\frac{kn_i}{2\pi} \right)^{d_i}}\, \left( \hat{\alpha}_{n_i}^\dag \right)^{d_{i}} | 0\rangle~, \quad\quad d_i , p \in \mathbb{N}~, 
\end{equation}
and their corresponding energy
\begin{equation}
E_{\{n_i,d_i\};m}  = k \, \upsilon \frac{m^2}{2} - \upsilon\left( \epsilon_0 + \frac{1}{24} \right) +  \upsilon \sum_{i=1}^p d_i n_i~.
\end{equation}
The degeneracy of those states for fixed $\sum_i n_i d_i = N$ is given by the integer partition number $p(N)$. We have allowed for an overall shift in the energy $\upsilon \epsilon_0$ to account for normal ordering ambiguities. The thermal partition function of the edge-mode theory is thus given by
\begin{equation}\label{zedge}
Z_{\text{edge}}[\beta] = \text{tr} \, e^{-\beta \hat{H}} =   {q^{-\epsilon_0}} \times \frac{\vartheta_3(0,q^{k/2})}{{\eta(q)}}~,
\end{equation}
where $q = e^{-\upsilon \beta}$, and $\vartheta_3(0,q)$ is the elliptic theta function
\begin{equation}
\vartheta_3 (z, q ) \equiv \sum_{n \in \mathbb{Z}}  q^{n^2} e^{2ni z} ~.
\end{equation}
We reproduce the above from the perspective of a Euclidean partition function in appendix \ref{Zthermal}.

\begin{figure}[H]
	\centering
\includegraphics[width=0.75\linewidth]{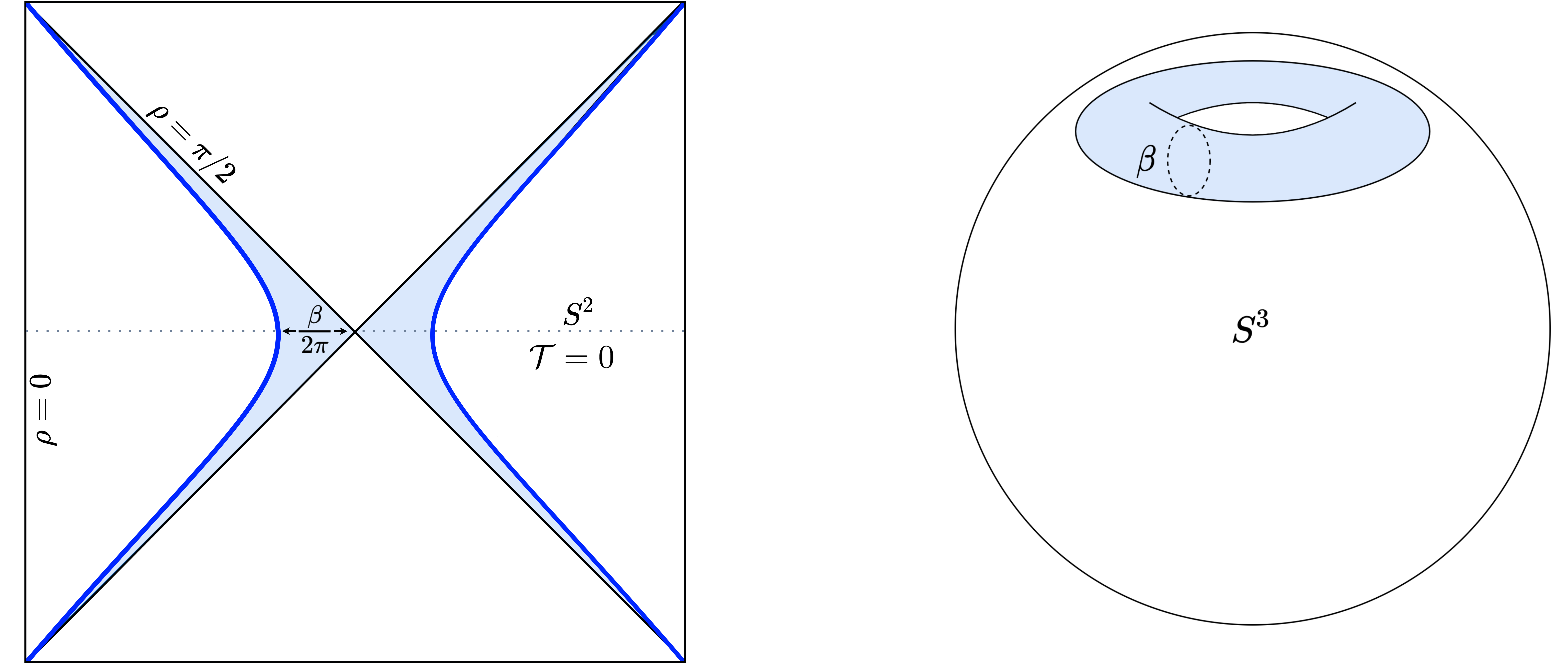}
	\caption{The Penrose diagram on the left shows the Lorentzian picture of the edge modes, which live on the thick, blue line at a distance $\beta/2\pi \ll \ell$ from the horizon. On the right is the Euclidean picture, where the edge modes live on the surface of a solid torus with thermal cycle $\beta$. Shrinking the thermal cycle is the same as taking the high temperature limit, ot taking the edge modes to live very close to the horizon.}
	\label{edgemodes}
\end{figure}

We now imagine that our edge-modes are located parametrically close to the dS$_3$ horizon. One might argue, in such a case, that the edge-mode theory should be placed at a parametrically high temperature (in units measured by the inertial clock at $\rho=0$). Let us discuss this from a Euclidean perspective. Recall that  Euclidean continuation of the de Sitter horizon becomes a circle in $S^3$. Removing a small region near the horizon corresponds to excising a thin solid torus $\bold{T}^2$ from the $S^3$, as shown in figure \ref{edgemodes}. The two cycles of the boundary of $\bold{T}^2$ correspond to a spatial cycle and a thermal cycle. As we take the region to vanishing size, the size of the thermal cycle shrinks to zero, which is the Euclidean picture of a high temperature limit. This is, to some extent, similar to the brick-wall regularisation considered by `t Hooft \cite{hooft1985quantum} (see also \cite{Das:2015oha,Geiller:2017xad,Wong:2017pdm}). Thus, to make contact with the dS$_3$ picture, we would like to study the edge-mode theory (\ref{edge}) at high temperature. 
In the high temperature limit, we can exploit the modular properties of $\vartheta(0,q^{k/2})$ and $\eta(q)$ to find
\begin{equation}\label{edgefinal}
\lim_{\beta\to0^+} \log Z_{\text{edge}}[\beta] = \frac{\pi^2}{6\upsilon\beta} -\frac{1}{2}  \log k \ldots
\end{equation}
The first term encodes the contribution from the high energy sector of the theory and is proportional to the temperature. The finite term is temperature independent, and moreover it is independent of our choice of $\upsilon$. In fact, it is equal to the three-sphere partition function (\ref{Zu1}). The entropy $S_{\text{e}}$ of the edge-modes at high temperature can similarly be computed and reads
\begin{equation}\label{Se}
S_{\text{e}} = \frac{\pi^2}{3\upsilon\beta} -\frac{1}{2}  \log k \ldots 
\end{equation}
Thus, we can relate the regularised expression for $Z_{U(1)_k}[S_3]$ (\ref{Zu1}) to the finite part of the entropy of the edge-mode theory in the high temperature limit. The divergent high temperature contribution is most naturally accommodated by a linearly divergent ultraviolet piece of the three-sphere partition function, indicating that
\begin{equation}
Z_{U(1)_k}[S_3] = \lim_{\beta\to0^+}  Z_{\text{edge}}[\beta]~,
\end{equation}
in a regularisation scheme where the linear divergence of $Z_{U(1)_k}[S_3]$ in (\ref{AbelianZ}) is tuned accordingly. We note that there are no cubic divergences appearing on either side. 

It has been argued \cite{kitaev2006topological,Levin:2006zz} that the temperature independent term in (\ref{edgefinal}) is a universal contribution to the entanglement entropy, independent of any of the detailed features associated to the cutoff surface. Here, in view of our discussion in section \ref{graventropy}, we can interpret $-\log \sqrt{k}$ as a contribution to the dS$_3$ horizon entropy. From this perspective, the fact that $-\log \sqrt{k}$ is negative is immaterial. 

\subsection{Edge-mode symmetries.} 

It is worth pausing momentarily here and noting that the edge-mode theory, which is nothing more than a chiral compact boson theory, has a large symmetry group containing an affine extension of the Virasoro algebra.\footnote{The dS$_3$/CFT$_2$ literature discusses a Virasoro symmetry associated to the future/past boundary \cite{strominger2001ds}. Here we see the appearance of a Virasoro symmetry within a single static patch.} The Virasoro generators are constructed in the usual way
\begin{equation}\label{virasoro}
\hat{L}_0 =  \frac{k}{2\pi}\sum_{n\in \mathbb{Z}^+} n^2 \hat{\alpha}^\dag_{n} \hat{\alpha}_{n}~, \quad\quad   \hat{L}_{m} = \frac{k}{2\pi}  \sum_{n=1}^\infty  (m+n)n \hat{\alpha}^\dagger_{n} \hat{\alpha}_{n+m} + \frac{k}{4\pi}\sum_{n=1}^{m-1}  (m-n)n\hat{\alpha}_{m-n} \hat{\alpha}_{n}~,  
\end{equation}
giving rise to the corresponding Virasoro algebra
\begin{equation} 
[\hat{L}_m, \hat{L}^\dag_n] = 
\begin{cases}
(m+n) \hat{L}_{m-n} + \frac{1}{12} m(m^2-1) \delta_{n,m}, \quad m \geq n~, \\
(m+n) \hat{L}^\dagger_{n-m} , \quad m < n ~. 
\end{cases}
\end{equation}
We notice that the edge-mode Hamiltonian in (\ref{eHam}) obeys $\hat{H}_0 = \upsilon \hat{L}_0$. We can further compute the commutation relations between the $\hat{L}_n$ and $\hat{\alpha}_n$ 
\begin{equation}
[\hat{L}_n,\hat{\alpha}_m] = -(n+m) \hat{\alpha}_{n+m}~, \quad\quad [\hat{L}_n,\hat{\alpha}^\dag_m] = \begin{cases}
		(m-n) \hat{\alpha}^\dag_{m-n}, \quad m>n\\
		(n-m) \hat{\alpha}_{n-m}, \quad m< n ~.
	\end{cases}
\end{equation}
The presence of the vertex operators (\ref{Psi}) allows the algebra to be extended beyond the Virasoro-$\mathfrak{u}(1)$ Kac-Moody algebra. The generators additionally include the vertex operators $\hat{\mathcal{J}}^{(k)}_\pm = : e^{\pm i {k} \hat{\zeta}}:$. Recalling that $k$ is even, it follows that the conformal dimension of the generators, $\Delta_\pm = k/2$, is also an integer.
The extended algebra has a finite number, $k$, of highest weight irreducible representations (see \cite{difrancesco} for a pedagogical discussion). We can organise the Hilbert space of the edge-mode theory in terms of these. The different representations are labelled by a parameter $s=0,1,\ldots,k-1$, denoting the representation carried by the Wilson line inserted in the interior of the disk \cite{witten1989quantum}. Their corresponding character is given by
\begin{equation} \label{u(1) character}
\chi_s(q)  = \frac{1}{\eta(q)} \sum_{m\in \mathbb{Z}} q^{\frac{k}{2} \left(m + \frac{s}{k}\right)^2}~, \quad\quad s = 0,1,\ldots,k-1~.
\end{equation}  
We observe that upon fixing $\epsilon_0 = 0$ and $q = e^{-v \beta}$, the above character for $s=0$ is equivalent to the thermal partition (\ref{zedge}). The modular $S$-matrix for the extended $\mathfrak{u}(1)$ character at level $k$ is given by (see for example \cite{Schellekens:1996tg,Dong:2008ft})
\begin{equation}\label{modsu1}
{\mathcal{S}_{m}}^{n} = \sqrt{\frac{1}{k}} e^{2\pi i m n/k}~, \quad\quad m,n = 0,1,\ldots,k-1~.
\end{equation}
Using that ${\mathcal{S}_{m}}^{n}$ relates the $\chi_s(q)$ under $\beta\to1/\beta$, we can re-derive the high temperature behaviour (\ref{edgefinal}). 

\subsubsection*{{\it Odd values of $k$ \& $\mathcal{N}=2$ supersymmetry, briefly}}
Before moving on to the non-Abelian case, we briefly comment that for odd  $k$ much of the above discussion carries through. The essential difference is that the generators $\hat{\mathcal{J}}^{(k)}_\pm = \, : e^{\pm i {k} \hat{\zeta}}:$ will carry half-integer conformal dimensions and obey fermionic anti-commutation relations. 

Let us focus on the case $k=3$, which is studied for example in \cite{Fendley:2006gr}. In this case the generators $\hat{\mathcal{J}}^{(3)}_\pm = :e^{\pm {3}i  \hat{\zeta}}:$ have weight $\Delta_\pm = 3/2$. Decomposing the generators into their respective Fourier modes $\hat{\mathcal{J}}^{(3)}_{\pm,n}$ on the cylinder,  one finds the anti-commutation relations
\begin{equation}
	\begin{split}
	&\{ \hat{\mathcal{J}}^{(3)}_{\pm,n}, \hat{\mathcal{J}}^{(3)}_{\pm,m} \} =  0~,\\
	&\{\hat{\mathcal{J}}_{+,n}^{(3)}, \hat{\mathcal{J}}_{-,m}^{(3)} \} = \frac{1}{2} \left(n^2 - \frac{1}{4} \right) \delta_{n+m,0} + 3 \hat{L}_{n+m } + \frac{3 i}{2} \left(m^2 - n^2 \right) \hat{\alpha}_{n + m}~.
	\end{split}
\end{equation} 
We can also compute the commutation relations between the $\hat{\mathcal{J}}^{(3)}_\pm$, the Virasoro generators (\ref{virasoro}), and the Kac-Moody generators (\ref{commutators}) leading to an $\mathcal{N}=2$ superconformal algebra.
This has been explored, for example, in \cite{Fendley:2006gr,Sagi:2016slk}.\footnote{The edge-mode theory of Chern-Simons theory with an $SU(2)$ gauge group at level $k=2$ also enjoys an $\mathcal{N}=2$ superconformal symmetry.} The operators $: e^{\pm  i  \hat{\zeta}}:$ correspond to insertions of fractional charge $\pm 1/3$ inside the spatial disk, and are primaries under the $\mathcal{N}=2$ superconformal algebra.

Supersymmetry is not often associated to physics in de Sitter space \cite{Pilch:1984aw}, but the usual arguments do not preclude the possibility of a supersymmetric edge-mode theory (or a superconformal theory more generally \cite{Anous:2014lia}).

\subsection{Comments on the non-Abelian case} \label{Comments on the non-Abelian case}

From the modular $S$-matrix (\ref{modsu1}) we observe that inserting a single Wilson line in the interior of the disk will not affect the constant part of the partition function because ${\mathcal{S}_{m}}^{0} = {\mathcal{S}_{0}}^{0}$. This is no longer true for the non-Abelian case. For instance, if we take Chern-Simons theory with gauge group $SU(2)$ at level $k$ the relevant modular $S$-matrix is (see for example \cite{difrancesco})
\begin{equation}\label{modsu2}
{\mathcal{S}_{m}}^{n} = \sqrt{\frac{2}{k+2}} \sin \left(\frac{(m+1)(n+1)\pi}{k+2} \right)~, \quad\quad m,n = 0,1,\ldots,k~.
\end{equation}
As mentioned earlier, ${\mathcal{S}_{0}}^{0}$ is equal to the regularised partition function $Z_{SU(2)_k}[S^3]$ for the choice of framing leading to vanishing phase. Explicitly,
\begin{equation}\label{zsu2s3}
Z_{SU(2)_k}[S^3] =  \sqrt{\frac{2}{k+2}} \sin \left(\frac{\pi}{k+2} \right)~.
\end{equation}
When considering the theory on the disk, the thermal partition function of the edge-mode theory is again given by a character of the $\widehat{\mathfrak{su}}(2)_k$ extended algebra. These are known from the rational CFT literature. For an insertion of the spin-$s/2$ integrable highest weight representation of level $k$ in the interior of the disk, we have (see for example \cite{difrancesco})
\begin{equation}\label{su2char}
Z_{\text{edge}}[\beta;s]  = \frac{\vartheta_{s +1}^{\, k+2} - \vartheta_{-s - 1}^{\, k+2}}{\vartheta_1^{\,2} - \vartheta_{-1}^{\,2}}~, \quad\quad s = 0,1, \ldots,k~,
\end{equation}
where the generalised $\vartheta$-functions are
\begin{equation}
\vartheta_s^{\, k }(q,z) \equiv \sum_{p \in \tfrac{s}{2k} + \mathbb{Z}} q^{kp^2}e^{2\pi i p k z}.
\end{equation}
As in the Abelian case, $q = e^{-\beta \upsilon}$ and $z=0$ in (\ref{su2char}) gives the thermal partition function of the edge-mode theory. The entropy in the high temperature limit of the edge-mode theory, which is now an $SU(2)_k$ chiral WZW model, will be given by
\begin{equation}\label{nonabelianedge}
	S^{(s)}_{\text{e}}  = \frac{3 k}{k+2}\, \frac{\pi^2}{3\beta v}  + \log \left(  \sqrt{\frac{2}{k+2}} \sin \left(\frac{(s+1)\pi}{k+2}\right) \right) \ldots
\end{equation}
The above is an increasing function of $s$ in the range $s=0,1,\ldots,k$.\footnote{It is also  customary to express the entropy in terms of the quantum dimensions $d^{(k)}_s$ of $SU(2)_k$. The numbers encode the multiplicity of various operators appearing upon fusion. They are given by
\begin{equation}
d^{(k)}_s = {\sin \left( \frac{\pi(s+1)}{k+2} \right)}{\sin^{-1} \left( \frac{\pi}{k+2} \right)}~, \quad\quad s = 0,1,2,\ldots,k~,
\end{equation}
such that the temperature independent part of the entropy is given by $S^{(s)}_{\text{e}} = -\frac{1}{2}\log \sum_{s} d^{(k)}_s d^{(k)}_s$.
In the large $k$ limit the quantum dimensions are approximately given by the usual degeneracy formula for $SU(2)$ spin-$s/2$ irreducible representations $d^{(k)}_s \approx s + 1$~.}
For vanishing $s$, the temperature independent piece of $S_{\text{e}}$ is the $S^3$ partition function (\ref{zsu2s3}) of Chern-Simons theory with $SU(2)$ gauge group at level $k$. For non-vanishing $s$, it is the $S^3$ partition function including the insertion of a closed unknotted Wilson loop carrying the spin-$s/2$ integrable representation of level $k$. The large $k$ expansion makes clear that in this case, horizon thermodynamics will receive contributions at all loop orders. In the weakly coupled limit, one finds
\begin{equation}
\lim_{k\to\infty} \exp {S^{(s)}_{\text{e}}}  \approx (1+s) \times \frac{1}{\text{vol} \, SU(2)} \left(\frac{4\pi^2}{k} \right)^{3/2}\times e^{\frac{\pi^2}{\beta v}}~, \quad s = 0,1,\ldots,k~.
\end{equation}
We see that the exponential of the entropy receives a multiplicative factor of $(1+s)$ in the weakly coupled limit. The factor $(1+s)$ is nothing more than the dimension of the Hilbert space attached to the puncture, and hence the entropy increases by this amount if the state is not measured with further precision.

Finally, we can also consider the phase of $Z_{SU(2)_k}[S^3]$ from the perspective of the edge-mode theory, again following \cite{witten1989quantum}. Usually, when we consider a two-dimensional conformal field theory on the torus, the partition function is invariant under shifts $\tau \to \tau+1$. This is simply the statement that the angular momentum is quantised. However, our edge-mode theory is chiral, and  transforms anomalously upon shifting $\tau\to\tau+1$. The integrable character of weight $h$ transforms as 
\begin{equation}
\chi_h(\tau+1) = e^{2\pi i (h-c_k/24)} \chi_h(\tau)~,
\end{equation}
where for the $SU(2)_k$ theory under consideration the weights of the spin-$s/2$ primaries are given by $h = s(s+2)/4(k+2)$ and $c_k = 3k/(k+2)$. In particular, under $\tau \to \tau+n$ the identity character transforms by a phase given by $e^{i n  c_k/24}$ where $n$ is an integer. This agrees with the set of admissible phases for $Z_{SU(2)_k}[S^3]$ stemming from the framing anomaly. 
\section{Complexified Abelian Chern-Simons: Lorentzian model}\label{complexCSL}

In this section we proceed to consider a complexified version of the Abelian Chern-Simons theory. This theory is introduced as a simple and calculable toy model with a complexified gauge group, a property common to three-dimensional gravity with Lorentzian signature and $\Lambda > 0$ expressed as a Chern-Simons theory.

\subsection{Lorentzian model}

The Lorentzian model is built from a complexified gauge field $\mathcal{A}_\mu = A_\mu + i B_\mu$, where $A_\mu$ and $B_\mu$ are real Abelian gauge fields. Our action is given by
\begin{equation}\label{complexSL}
S_L[\mathcal{A}_\mu] = \frac{k + i \lambda}{8\pi} \int_{\mathcal{M}} d^3 x  \, \varepsilon^{\mu\nu\rho} \mathcal{A}_\mu \partial_\nu \mathcal{A}_\rho +  \frac{k - i \lambda}{8\pi} \int_{\mathcal{M}} d^3 x  \, \varepsilon^{\mu\nu\rho} \bar{\mathcal{A}}_\mu \partial_\nu \bar{\mathcal{A}}_\rho~,
\end{equation}
and is real-valued. The parameters $k$ and $\lambda$ are taken to be real-valued. We denote the Lie algebra of our theory by $\mathfrak{u}_{\mathbb{C}}(1)$. The gauge transformations act in the following way
\begin{equation}
\mathcal{A}_\mu \to \mathcal{A}_\mu + \partial_\mu \vartheta~,
\end{equation}
where the gauge parameter $\vartheta$ is a complex-valued function.
In terms of $A_\mu$ and $B_\mu$ the action is given by
\begin{equation}\label{ABaction}
S_L[\mathcal{A}_\mu] = \frac{k}{4\pi} \int_{\mathcal{M}} d^3 x  \, \varepsilon^{\mu\nu\rho}  \left( A_\mu \partial_\nu A_\rho- B_\mu \partial_\nu B_\rho   \right) - \frac{\lambda}{4\pi}  \int_{\mathcal{M}} d^3 x  \, \varepsilon^{\mu\nu\rho}   \left( A_\mu \partial_\nu B_\rho + B_\mu \partial_\nu A_\rho \right)~.
\end{equation}
At $\lambda=0$, the real and imaginary parts of the gauge field $\mathcal{A}$ decouple. Setting either $\lambda=0$ or $k=0$ reduces the model to an Abelian BF model with $U(1)$ gauge group (see \cite{Birmingham:1991ty,Cattaneo:1995tw} for an overview). In addition to being real-valued, we will impose that $k$ take integer values upon quantisation of the theory. This follows from taking $A_\mu$ as the generator of a compact $\mathfrak{u}(1)$ gauge algebra. The parameter $\lambda$ is not constrained to be an integer. Unless otherwise specified, we will take $k \in \mathbb{Z}^+$ and $\lambda \in\mathbb{R}/\{0\}$ in what follows. 

The classical equations of motion for (\ref{ABaction}) are given by
\begin{equation}
\varepsilon^{\mu\nu\rho} \partial_\nu A_\rho =  \varepsilon^{\mu\nu\rho} \partial_\nu B_\rho= 0~.
\end{equation}
They are satisfied when the field strengths associated to $A_\mu$ and $B_\mu$ vanish. The classical solution space is the space of flat connections modulo gauge transformations. Assuming the theory resides on a three-manifold with topology $\mathbb{R} \times S^2$, we can canonically quantise the theory in the $A_t = B_t = 0$ gauge. States must then satisfy the constraints
\begin{equation}\label{complexconstraint}
F^{A}_{ij} =   F^{B}_{ij} = 0~, \quad\quad i,j \in S^2~.
\end{equation}
The equal time commutation relations are given by
\begin{equation}\label{complexcomm}
[A_i(\bold{x}), A_j (\bold{y}) ] = - [B_i(\bold{x}), B_j (\bold{y}) ] = \frac{2\pi i}{k} \varepsilon_{ij} \delta (\bold{x} - \bold{y} ) ~.
\end{equation}
It follows that on a spatial $S^2$ the Hilbert space has a unique state.

As a final remark before embarking onto the edge-mode theory, it is worth noting that the volume of $U_{\mathbb{C}}(1)$ is infinite, making it difficult to interpret the perturbative formula (\ref{CSpert}) for the $S^3$ partition function of the Lorentzian model.

\subsection{Lorentzian edge-mode theory \& quantisation}

Following the discussion in section \ref{edgemodeu1}, we can construct an edge-mode theory at the $S^1$ boundary of the spatial disk. As before, we consider fixing the following gauge\footnote{More general boundary conditions can also be considered, and it may be interesting to do so.} 
\begin{equation}
\mathcal{A}_t - \upsilon \mathcal{A}_\varphi = 0~,
\end{equation}
where $\upsilon = \upsilon_{re} + i \upsilon_{im}$ is now a complex parameter. The gauge constraint that follows from imposing the above condition is
\begin{equation}
\mathcal{A}_i (t,\rho,\varphi) = i e^{i \, \Xi(t,\rho,\varphi)} \, \partial_i \, e^{-i \,  \Xi(t,\rho,\varphi)}~, \quad\quad i \in \{ \rho, \varphi \}~,
\end{equation}
where now the boundary value of $\Xi(t,\rho,\varphi)|_{\rho=\pi/2}$ is denoted by $\zeta = \zeta_{re} + i \zeta_{im}$ and has compact real part. The action governing our edge-mode theory is given by
\begin{equation}\label{SLedge}
S_{\text{edge}} = \frac{k+i \lambda}{8\pi} \int d t d\varphi  \left( \partial_t \zeta - \upsilon \partial_\varphi \zeta \right) \partial_\varphi \zeta +  \frac{k-i \lambda}{8\pi} \int d t d\varphi  \left( \partial_t \bar{\zeta}  - \bar{\upsilon} \partial_\varphi \bar{\zeta}  \right) \partial_\varphi \bar{\zeta}~.
\end{equation}
Once again, the classical solutions are given by complexified chiral excitations of both the real and imaginary parts of $\zeta$. We find 
\begin{equation}
\zeta = f(\varphi+ \upsilon t) + m \varphi + g(t)~,
\end{equation}
where $f(z)$ is a complex valued function. We note that $\zeta_{im}=  (\zeta-\bar{\zeta})/2i$ does not contain winding modes around the $S^1$. The classical Hamiltonian is given by
\begin{equation}
\mathcal{Q}_t = \frac{(k+i \lambda) \upsilon}{8\pi} \int d\varphi  \left( \partial_\varphi \zeta  \right)^2 +  \frac{(k- i \lambda)\bar{\upsilon}}{8\pi} \int d\varphi  \left( \partial_\varphi \bar{\zeta}  \right)^2~,
\end{equation}
which generates time translations. We note that $\mathcal{Q}_t$ is real. The generator of $\varphi$-translations is given by
\begin{equation}
\mathcal{Q}_\varphi = \frac{(k+i \lambda) }{8\pi} \int d\varphi   \partial_\varphi \zeta  \partial_t \zeta + \frac{(k- i \lambda)}{8\pi} \int d\varphi   \partial_\varphi \bar{\zeta} \partial_t \bar{\zeta}~.
\end{equation}
On-shell $\mathcal{Q}_t$ and $\mathcal{Q}_\varphi$ are equivalent within the sector with vanishing winding number. The classical theory (\ref{SLedge}) is also invariant under shifts $\zeta \to \zeta + \delta$ of the field $\zeta$ by some $\delta \in \mathbb{C}$. The generators of the shift symmetries are
\begin{equation}\label{complexQ}
\mathcal{Q}_\zeta = \frac{1}{2\pi} \int d\varphi \partial_\varphi \zeta~, \quad\quad \mathcal{Q}_{\bar{\zeta}} = \frac{1}{2\pi} \int d\varphi \partial_\varphi \bar{\zeta}~.
\end{equation}
\textbf{Quantisation.} In order to quantise the theory, it is convenient to express $\zeta$ in terms of Fourier modes. For the non-winding mode sector, we have
\begin{equation}
\zeta(t,\varphi) = \frac{1}{\sqrt{2\pi}} \sum_{n\in\mathbb{Z}/\{0\}} \alpha_n(t) e^{i n \varphi}~.
\end{equation}
The $\alpha_n(t) \equiv a_n(t) + i b_n(t)$ are independent complex functions of $t$ for all $n$, and we take $a_n(t)$ and $b_n(t)$ to be real-valued. To quantise the theory, it is convenient to follow the procedure outlined in \cite{Dunne:1992ew}. For a given $n$, we can define the vector with components $\xi^i_n$ by $\xi_n \equiv (a_{-n},b_{-n},a_n,b_n)$. One finds the commutator
\begin{equation}
[\hat{\xi}^i_n,\hat{\xi}^j_n] = \frac{i}{n}  M^{ij}~, 
\end{equation} 
with
\begin{equation}
M^{-1} = \frac{2\pi}{k^2 + \lambda^2} \left(
\begin{array}{cc}
 0 & -1 \\
 1 & 0 \\
\end{array}
\right) \otimes \left(
\begin{array}{cc}
 \lambda  & k  \\
 k  & -\lambda  \\
\end{array}
\right)~.
\end{equation}
The eigenvalues of $M^{-1}$ are $\pm 2\pi i/ \sqrt{k^2+\lambda^2}$, each doubly degenerate. In terms of the $\hat{\xi}_n$, the quantum Hamiltonian for the vanishing winding mode sector is given by
\begin{equation}
\hat{\mathcal{Q}}_t = \frac{1}{4\pi} \sum_{n\in\mathbb{Z}/\{0\}} n^2 \left(\omega \left(\hat{\xi}^3_n \hat{\xi}_n^1 - \hat{\xi}^4_n \hat{\xi}_n^2 \right) - \psi \left(\hat{\xi}^3_n \hat{\xi}^2_n + \hat{\xi}^4_n \hat{\xi}^1_n \right) \right)~.
\end{equation}
Here we have defined $\omega \equiv k \upsilon_{re} - \lambda \upsilon_{im}$ and $\psi \equiv  \lambda \upsilon_{re} + k \upsilon_{im}$. We can  construct operators obeying the more standard creation/annihilation algebra. We find that
\begin{equation}
	\begin{split}
		\hat{A}_n^{\pm} & \equiv \hat{\xi}_n^4 \pm i \hat{\xi}_n^3 \pm \frac{(ik\mp  \lambda)  }{\sqrt{k^2 + \lambda^2 }} \left(\hat{\xi}_n^1 \pm i \hat{\xi}_n^2 \right)~,\\   
		\hat{B}_n^{\pm} & \equiv \hat{\xi}_n^4 \mp i \hat{\xi}_n^3 \pm \frac{(ik \pm \lambda)}{\sqrt{k^2 + \lambda^2}} \left(\hat{\xi}_n^1 \mp i  \hat{\xi}_n^2\right)~,
	\end{split}
\end{equation}
obey the commutation relations
\begin{equation}\label{complexAAdag}
	\begin{split}
		[\hat{A}_n^+, \hat{A}_m^-] = [\hat{B}_n^+, \hat{B}_m^-] = \frac{8\pi}{n\sqrt{k^2 + \lambda^2}} \delta_{n,m}~,
	\end{split}
\end{equation}
with all others vanishing. The Hamiltonian can then be put into the following form

\begin{equation}\label{LedgeH}
\hat{\mathcal{Q}}_t = \frac{\sqrt{k^2+\lambda^2}}{16\pi} \sum_{n\in\mathbb{Z}/\{0\}} n^2 \left( \upsilon_{re} \left( \hat{A}_n^+ \hat{A}_n^- - \hat{B}_n^+ \hat{B}_n^- \right) + i  \upsilon_{im} \left( \hat{A}_n^+ \hat{B}_n^+ - \hat{A}_n^- \hat{B}_n^-  \right) \right)~.
\end{equation}
Consequently, uncovering the spectrum of $\hat{\mathcal{Q}}_t$ reduces to a two-site hopping type problem, where the sites are labelled by ${A}$ and $B$.

\subsection{Lorentzian edge-mode spectrum}

At this stage, we are confronted with diagonalising the Lorentzian edge-mode Hamiltonian (\ref{LedgeH}). Since each Fourier mode decouples, it is sufficient to discuss the problem for a single mode number $n$. It is convenient to represent the creation and annihilation operators satisfying (\ref{complexAAdag}) in the following way
\begin{equation}
\hat{A}^\pm_n = \left( \frac{4\pi}{n\sqrt{k^2 + \lambda^2}}  \right)^{1/2} \left( x_n \pm \frac{d}{dx_n} \right)~, \quad \hat{B}^\pm_n =  \left( \frac{4\pi}{n\sqrt{k^2 + \lambda^2}}  \right)^{1/2} \left( y_n \pm \frac{d}{dy_n} \right)~.
\end{equation}
As detailed in appendix \ref{LedgeApp}, it is straightforward to put the single mode Hamiltonian $\hat{\mathcal{Q}}_t^{(n)}$ problem into the Schr\"odinger problem
\begin{equation}\label{Qn}
 \frac{\text{sign} \, w_n}{2}\left(- \frac{d^2}{dw_n^2 } +w_n^2 - \frac{1/4+p_n^2}{w_n^2}\right) \psi_n = \frac{1}{\upsilon_{re} } \left( \frac{E_n}{n} +  \upsilon_{im} p_n\right) \psi_n~,
\end{equation}
where $x_n = w_n \cosh t_n $, $y_n = w_n\sinh t_n $ and $p_n \in \mathbb{R}$ is the Fourier momentum associated to the $t_n \in \mathbb{R}$ coordinate. The $E_n$ are the eigenvalues of $\hat{\mathcal{Q}}_t^{(n)}$. The operator on the left-hand side of (\ref{Qn}) is precisely that of a conformal quantum mechanics as studied for example in \cite{de1976conformal,Anous:2020nxu,andrzejewski2011geometry, andrzejewski2016quantum}, whose $sl(2,\mathbb{R})$ generators are given by
\begin{equation}\label{sl2rgen}
\hat{K}_n = \frac{\text{sign} \, w_n}{2} w_n^2~, \quad \hat{D}_n = -  \frac{ i}{2} \left[ w_n \frac{d}{dw_n} + \frac{1}{2} \right]~, \quad \hat{H}_n = -\frac{\text{sign} \, w_n}{2} \left[\frac{d^2}{dw_n^2} + \frac{1/4+p^2}{w_n^2} \right]~,
\end{equation}
satisfying
\begin{equation}
[\hat{D}_n,\hat{H}_n] = i\hat{H}_n~, \quad\quad [\hat{D}_n,\hat{K}_n]=-i\hat{K}_n~, \quad\quad [\hat{K}_n,\hat{H}_n] = 2i\hat{D}_n~. 
\end{equation}
The generators (\ref{sl2rgen}) furnish the principal series representation with weight $\Delta = 1/2+ i p_n/2$. 
It is known that the spectrum of $\hat{H}_n+\hat{K}_n$ is given by the even integers \cite{Anous:2020nxu,andrzejewski2011geometry, andrzejewski2016quantum}, and is thus unbounded. Consequently, the single mode energy spectrum is given by
\begin{equation} \label{Loreenergies}
E_n(m,p_n) = 2\upsilon_{{re}} n m - \upsilon_{{im}} n p_n~, \quad\quad m \in \mathbb{Z}~, \quad p_n \in \mathbb{R}~.
\end{equation}
The unboundedness and continuity of the edge-mode Hamiltonian is a reflection of the fact that the Lorentzian Chern-Simons theory has a complexified gauge group. It also implies that the thermal partition function of the edge-mode theory is no longer a sensible quantity to compute. (Sensible quantities to compute might involve \cite{Carlip:1994gy,Banados:1996ad} imposing additional constraints on the state-space or modifying the boundary conditions on the Chern-Simons gauge field at the boundary of the disk.) To sharpen this issue, we consider the edge-mode theory on a torus.

\subsection{Lorentzian model in Euclidean signature?}\label{complexL}

The Euclidean continuation of the complexified edge-mode theory (\ref{SLedge}) can be achieved by taking $t = -i\tau$. If we are to place the system at a finite inverse temperature $\beta$, we further impose that $\tau \sim \tau + \beta$, such that the theory resides on a torus. Upon continuing to Euclidean time, we end up with the Euclidean edge-mode action
\begin{equation}\label{SLEedge}
S^{(E)}_{\text{edge}} = \frac{k+i \lambda}{8\pi} \int d \tau d\varphi  \left(- i \partial_\tau \zeta + \upsilon \partial_\varphi \zeta \right) \partial_\varphi \zeta +  \frac{k-i \lambda}{8\pi} \int d \tau d\varphi  \left( - i  \partial_\tau \bar{\zeta}  + \bar{\upsilon} \partial_\varphi \bar{\zeta}  \right) \partial_\varphi \bar{\zeta}~.
\end{equation}
It is convenient to further express the action in terms of the modes on the torus
\begin{equation}\label{Fedge}
\zeta(\tau,\varphi) = \frac{1}{\sqrt{2\pi}} \sum_{(m,n) \in \mathbb{Z}^2} e^{2\pi i m  \tau/\beta +  i n \varphi} \zeta_{m,n}~, \quad \bar{\zeta
}(\tau,\varphi) = \frac{1}{\sqrt{2\pi}} \sum_{(m,n) \in \mathbb{Z}^2} e^{-2\pi m i \tau/\beta -  i n \varphi} \bar{\zeta}_{m,n}~,
\end{equation}
such that
\begin{multline}\label{SLEedgemodes}
S^{(E)}_{\text{edge}} =\sum_{(m,n) \in \mathbb{Z}^2}  \frac{k+i \lambda}{8\pi}\left( -{2\pi i m}n+\beta \upsilon n^2  \right) \zeta_{m,n} \zeta_{-m,-n}+ \frac{k-i \lambda}{8\pi} \left( - {2\pi i m} n+ {\beta}\bar{\upsilon} n^2  \right) \bar{\zeta}_{m,n} \bar{\zeta}_{-m,-n}~.
\end{multline}
Here $\zeta_{m,n} \in \mathbb{C}$ and $\bar{\zeta}_{m,n}$ is the complex conjugate of $\zeta_{m,n}$. As discussed in appendix \ref{Zthermal}, in the Abelian case a Euclidean continuation that preserves the reality conditions of the Chern-Simons gauge field would further require the continuation $\upsilon = -i \upsilon_E$. In the complexified case, this is no longer necessitated as $\upsilon$ is allowed to take complex values. Instead, given the unboundedness of the spectrum we must confront a Gaussian unsuppressed Euclidean path-integral
\begin{equation} \label{deformed}
\mathcal{Z}_{\text{edge}} [\beta] = \int \mathcal{D}\zeta \mathcal{D} \bar{\zeta} \, e^{-S^{(E)}_{\text{edge}}[\zeta,\bar{\zeta}]}~.
\end{equation}
In order to render $\mathcal{Z}_{\text{edge}} [\beta]$ better defined, we can complexify the contour of path-integration \cite{Witten:2010cx}. For simplicity, we take $\upsilon_{re}  > 0$ and $\lambda=0$ while recalling $k \in \mathbb{Z}^+$. It is then clear that taking a contour where $\zeta(\tau,\varphi)$ and $\bar{\zeta}(\tau,\varphi)$ are independent real fields will render $\mathcal{Z}_{\text{edge}} [\beta]$ well-defined.\footnote{Expressing $\mathcal{Z}_{\text{edge}} [\beta]$ in terms of the Fourier modes (\ref{Fedge}), we can view the problem as an infinite-dimensional version of the complex integral
\begin{equation}
\mathcal{I}_\sigma = \int_{\mathbb{C}\times \mathbb{C}} dz d\bar{z} dw d\bar{w} \, e^{- \sigma z w -  {\sigma} \bar{z} \bar{w}}~,
\end{equation}
where $\sigma \in \mathbb{C}$ with positive real part. To define $\mathcal{I}_\sigma$ we first take $z$ and $\bar{z}$ as well as $w$ and $\bar{w}$ to be independent complex variables, and then integrate over the contour $w= \bar{z}$ and $\bar{w} = z$. In this sense, we have that $\mathcal{I}_\sigma = {\pi^2}/{{\sigma^2}}$.}
Upon continuing the contour of $\zeta$ and $\bar{\zeta}$, the symmetry group of the edge-mode theory becomes $U(1)\times U(1)$. As we will see in the following section, the price to pay in curing the unboundedness of the Hamiltonian is unitarity of the edge-mode theory. For $\lambda \neq 0$, we must consider a more elaborate contour to render $\mathcal{Z}_{\text{edge}} [\beta]$ better defined. This is discussed in the next section where we view the $\lambda \neq 0$ case as a continuation of the $\lambda=0$ case with $k \in \mathbb{Z}^+$.

From the perspective of the original Chern-Simons gauge theory (\ref{complexSL}) with complexified gauge group $U_{\mathbb{C}}(1)$, the analytic continuation rendering $\zeta(\tau,\varphi)$ and $\bar{\zeta}(\tau,\varphi)$ as independent real fields can be achieved by continuing the complexified Chern-Simons gauge fields $\mathcal{A}_\mu$ and $\bar{\mathcal{A}}_\mu$ to two independent real-valued $U(1)$ gauge fields, whilst maintaining a complexified level.

\section{Complexified Abelian Chern-Simons: Euclidean model}\label{complexCS}

In this section we proceed to consider a different complexified version of the Abelian Chern-Simons theory. This theory is introduced as a simple and calculable toy model with a complexified level, a property common to three-dimensional gravity with Euclidean signature and $\Lambda > 0$ viewed as a Chern-Simons theory.
 
\subsection{Euclidean model on $S^3$}

The action for the Euclidean model is given by
\begin{equation}\label{EucU1}
S_E[{A}^\pm_\mu] = \frac{\kappa + i \gamma}{4\pi} \int_{\mathcal{M}} d^3 x  \, \varepsilon^{\mu\nu\rho} {A}^+_\mu \partial_\nu {A}^+_\rho +  \frac{\kappa - i \gamma}{4\pi} \int_{\mathcal{M}} d^3 x  \, \varepsilon^{\mu\nu\rho} {A}^-_\mu \partial_\nu {A}^-_\rho~,
\end{equation}
where now ${A}^+$ and ${A}^-$ are two real-valued $\mathfrak{u}(1)$ gauge fields. We assume that both $U(1)$ groups are compact and take the parameter $\kappa \in \mathbb{Z}^+$. There is no restriction on $\gamma$ so we take $\gamma \in \mathbb{R}^+$. 
The $S^3$ partition function must be carefully defined since $i S_E[{A}^\pm_\mu]$ is no longer purely oscillatory but rather an unbounded functional rendering the path integral over $e^{i S_E}$ problematic. To make sense of it we can again consider a complexification of the original path-integration contour. On $S^3$ the only $U(1)$ flat connection (modulo gauge transformations) is the trivial one. Moreover, given that the action is quadratic in the fields the choice of contour can be reduced to a problem of Gaussian integration. The Gaussian integral we are interested in is of the form
\begin{equation}
\mathcal{I}_\sigma = \int_{\mathcal{C}} d x  \, e^{-\sigma x^2/2}~,
\end{equation}
with $\sigma = e^{i \vartheta}|\sigma| \in \mathbb{C}$. We take $\mathcal{C}$ to be along a path $e^{i\varphi} \tilde{x}$ with $\tilde{x} \in \mathbb{R}$ and $\vartheta+2\varphi \in (-\pi/2,\pi/2)$ such that
\begin{equation}
\mathcal{I}_\sigma = \sqrt{\frac{2\pi}{|\sigma|}} e^{-i\vartheta/2}~.
\end{equation}
We further take that $\mathcal{I}_{\bar{\sigma}} \equiv \left( \mathcal{I}_{\sigma} \right)^*$ such that the corresponding contour is $e^{-i\varphi} \tilde{x}$, and the product $\mathcal{I}_{\bar{\sigma}} \mathcal{I}_{\sigma}$ is positive and real. Applying the same reasoning to the Chern-Simons case, we arrive at
\begin{equation}\label{complexS3}
Z_{\kappa,\gamma}[S^3] = \left(\frac{1}{\text{vol} \, U(1)} \right)^2 \times \left| \frac{4\pi^2}{\kappa+i\gamma} \right| = \frac{1}{\sqrt{\kappa^2+\gamma^2}}~.
\end{equation}
The above result is invariant under complex conjugation, i.e. $\gamma\to-\gamma$. For $\gamma=0$ we retrieve the standard result (\ref{Zu1}) with a choice of framing leading to a vanishing phase. For $\kappa = 0$, the result makes sense for $\gamma \in \mathbb{R}^+$ but becomes non-analytic if one tries to continue to the whole complex-$\gamma$ plane.\footnote{Although our treatment leads to a real valued $Z_{\kappa,\gamma}[S^3]$, it seems feasible that a more general regularisation can lead to an overall phase that depends on the choice of framing of $S^3$. It would be interesting to explore this both for the Abelian case and the non-Abelian extension discussed in section \ref{EuclideanGrav}.}  

\subsection{Euclidean edge-mode theory on the torus}

Since $A^\pm$ are now compact $\mathfrak{u}(1)$ gauge fields, the boundary chiral bosons, which we denote as $\zeta^\pm$, are now compact. Following the procedure outlined in previous sections, and imposing the boundary condition 
\begin{equation}
A^\pm_\tau - \upsilon^\pm A^\pm_\varphi |_{\partial \mathcal{M}} = 0~,
\end{equation}
with $ \upsilon^\pm \in \mathbb{R}$, we end up with an edge-mode theory governed by the action
\begin{equation}
S_{\text{edge}} = \frac{\kappa+i\gamma}{4\pi} \int d \tau d\varphi \, \partial_\varphi \zeta^+  \left( \partial_\tau  - \upsilon^+ \partial_\varphi  \right) \zeta^+ +  \frac{\kappa-i\gamma}{4\pi} \int d \tau d\varphi \, \partial_\varphi \zeta^-  \left( \partial_\tau  - \upsilon^- \partial_\varphi  \right) \zeta^-~.
\end{equation}
The above theory, which we take to be in Euclidean signature, is non-unitary. Consequently, a Hilbert space interpretation of the theory is unclear. Nevertheless, we can try to compute the Euclidean torus partition function of $S_{\text{edge}}$. We take the periodicity of Euclidean time to be $\tau \sim \tau+\beta$, and we recall that $\varphi \sim \varphi+2\pi$. 
It is convenient to express the non-winding sector in terms of their respective Fourier modes, namely
\begin{equation}
\zeta^\pm(\tau,\varphi) = \frac{1}{\sqrt{2\pi}} \sum_{(m, n) \in \mathbb{Z}^2} \zeta^\pm_{m,n} e^{  2\pi i m \tau/ \beta +  i n \varphi}~.
\end{equation}
The reality condition is $\zeta^\pm_{-m,-n} = \left( \zeta^\pm_{m,n}\right)^*$. The edge-mode action now reads
\begin{equation}\label{SEedge}
S_{\text{edge}} = \sum_{(m,n)\in \mathbb{Z}^2} \frac{\kappa+i\gamma}{4\pi}  \left( {2\pi m n} -\beta \upsilon^+  n^2\right) |\zeta^+_{m,n}|^2 + \frac{\kappa-i\gamma}{4\pi}  \left( {2\pi m n} - \beta \upsilon^-  n^2\right) |\zeta^-_{m,n}|^2~,
\end{equation}
and the path integral becomes an integral over all the $\zeta^\pm_{m,n}$. We can compare the edge-mode action $i S_{\text{edge}}$ to the one stemming from Lorentzian edge-mode theory, i.e. $-S^{(E)}_{\text{edge}}$ in (\ref{SLEedgemodes}), discussed in the previous section. If we map $ (k,\lambda) = (2\kappa,2\gamma)$ and $(\upsilon,\bar{\upsilon}) = i (\upsilon^+,\upsilon^-)$, it follows that the deformed contour discussed below (\ref{deformed}) indeed leads to the Euclidean edge-mode theory (\ref{SEedge}). 

In order to render the path integral over $i S_{\text{edge}}$ convergent, we must again alter the integration contour $\mathcal{C}$ of $\zeta_{n,m}^\pm$. The path integration over the non-winding mode sector of $\zeta^\pm$ yields a functional determinant proportional to the Dedekind $\eta$-function (see appendix \ref{Zthermal}). The sum over the winding sector must be similarly defined. The winding modes are given by $\zeta^\pm(t,\varphi) = m_\pm \varphi$ where $m_\pm$ are integers.\footnote{Given the periodicity of $\zeta^\pm$, one is also tempted to consider the vertex operators $\hat{\mathcal{O}}^\pm_n = :e^{i n \hat{\zeta}^\pm}:$ carrying $U(1)$ charge (\ref{U1charge}) $\mathcal{Q}^\pm_n = n/(\kappa\pm i\gamma)$. Although $\mathcal{O}^\pm_n$ bear some similarity to the anyonic operators (\ref{Psi}) of ordinary Abelian Chern-Simons theory, their physical meaning is more obscure at complex level. For instance, in the ordinary Abelian case $k \in \mathbb{Z}^+$ corresponds to the number of anyonic species, something hard to understand when the level is complex.} Consequently, we must perform the sum
\begin{equation}
\mathcal{S}_{\text{w}} = \sum_{m_\pm \in \mathbb{Z}} e^{-i \frac{ \kappa + i \gamma}{2} \upsilon^+ \beta m^2_+  - i \frac{ \kappa - i \gamma}{2}  \upsilon^- \beta m^2_-} =  \vartheta_3\left(0,q_+^{( \kappa+ i \gamma)/2}\right)  \vartheta_3 \left(0,q_-^{( \kappa- i \gamma)/2}\right) ~.
\end{equation}
The above sum exists provided $- ( { \kappa \pm  i\gamma} )\beta \upsilon^\pm/2 \pi$ is in the upper-half plane. In the last equality, we have defined ${q_{\pm} \equiv e^{-i \upsilon^{\pm}\beta}}$. 
Putting it all together, upon path integrating over $e^{iS_{\text{edge}}}$ we end up with the torus partition function
\begin{equation}
Z_{\text{edge}}\left[\beta\right] = \frac{\vartheta_3\left(0,q_+^{( \kappa+ i \gamma)/2}\right)  \vartheta_3 \left(0,q_-^{( \kappa- i \gamma)/2}\right)}{\eta(q_+) \eta(q_-)}~.
\end{equation}
To render the Dedekind $\eta$-functions well-defined, we take $\upsilon^\pm$ to have a small negative imaginary part. Taking the $\beta\to 0^+$ limit we have 
\begin{equation}
\lim_{\beta \rightarrow 0^+}\log Z_{\text{edge}}\left[\beta\right] = \frac{\pi^2 }{6\beta i \upsilon^+} + \frac{\pi^2 }{6\beta i \upsilon^-} - \log |\kappa+i\gamma| \ldots
\end{equation}
The finite part agrees with the $S^3$ partition function (\ref{complexS3}). We further note that the above expression can be obtained from the Abelian edge-mode expression (\ref{edgefinal}) by a simple analytic continuation.

\section{Remarks on gravitation in three-dimensions}\label{oneloop3dgrav}

The classical action of general relativity in three-dimensions can be expressed as a Chern-Simons theory \cite{Achucarro:1987vz,witten19882+}. The gauge group depends on the signature and sign of the cosmological constant $\Lambda$. We restrict to $\Lambda = +1/\ell^2 >0$. In Euclidean signature the gauge group is $SU(2)\times SU(2)$ and the level is complex. In Lorentzian signature the gauge group becomes $SL(2,\mathbb{C})$, which can be viewed as the complexification of $SU(2)$ (or $SL(2,\mathbb{R})$). These Chern-Simons theories are natural non-Abelian extensions of the complexified Abelian theories explored in sections \ref{complexCSL} and \ref{complexCS}. The purpose of this section is to comment briefly on how the properties of the complexified Abelian theories generalise, leaving a more detailed analysis to future work. 

\subsection{Euclidean signature}\label{EuclideanGrav}

We begin by considering the Euclidean theory whose action is given by $S_E = S^+_E + S^-_E$, where
\begin{equation}\label{3dgravE}
S^\pm_E[{A}^\pm_\mu] = \frac{\kappa \pm i \gamma}{4\pi}  \int_{\mathcal{M}} d^3 x  \, \varepsilon^{\mu\nu\rho} \, \text{Tr}\left({A}^\pm_\mu \partial_\nu {A}^\pm_\rho + \frac{2}{3} A^\pm_\mu A^\pm_\nu A^\pm_\rho \right)~.
\end{equation}
The $A^\pm_\mu$ are $SU(2)$ gauge fields. In terms of the vielbein $e^i_\mu$ and spin-connection $\omega_{ij,\mu} = {\epsilon_{ijk}} {\omega^{k}}_\mu$,  where $\epsilon_{ijk}$ is the Levi-Civita symbol, one has
\begin{equation}
A^\pm_\mu =  \left( \omega_\mu^i \pm e^i_\mu\right) T_i~, \quad\quad ~T_i = \frac{1}{2} i \sigma_i~,
\end{equation}
where the $\sigma_i$ are the Pauli matrices. Substituting $A^\pm$ into (\ref{3dgravE}) leads to Einstein gravity in three-dimensions plus a parity-odd Chern-Simons type term for the spin connection. The imaginary part of the level, $\gamma \in \mathbb{R}^+$, is given by $\gamma = \ell/4 G$, while $\kappa \in \mathbb{Z}$ is the coupling of a parity-odd gravitational Chern-Simons term. 

The equations of motion stemming from (\ref{3dgravE}) are the flat connection equations, and the solutions are the space of flat connections modulo gauge transformations. For standard $SU(2)$ Chern-Simons theory on an $S^3$ we would discard the presence of multiple possible saddles due to the absence of non-trivial flat connections. However, when the Chern-Simons coupling is taken away from the integers one must exercise further caution. In particular, gauge transformations changing the winding number will no longer be trivial. So, the space of flat connections with differing winding number have different on-shell actions. Of these, most will not have a clear interpretation from the perspective of the gravitational theory since they will lead to non-invertible or otherwise non-standard vielbeins. This is a feature we did not have to confront in the Abelian toy models of the previous sections. We consider perturbative effects around the flat-connection $A^+ = g^{-1} dg$ and $A^- = 0$, where $g \in SU(2)$. Recalling that the group geometry of $SU(2)$ under the Haar metric is the three-sphere, this flat connection corresponds to the round $S^3$ in the gravitational theory as can be explicitly checked. Following reasoning analogous to the Euclidean $U(1)$ model with complexified levels, it has been argued that \cite{Gukov:2016njj,Anninos:2020hfj}
\begin{equation} \label{subleadingterms}
\mathcal{Z}_{\text{grav}} = e^{2\pi \gamma} \times e^{ i \varphi_{\text{grav}}} \left| \sqrt{\frac{2}{\kappa+i \gamma+2}} \sin {\frac{\pi}{\kappa+i \gamma + 2}} \right|^2~,
\end{equation}
to all orders in a large-$\gamma$ perturbative expansion, and this expansion reproduces (\ref{logZgrav}).
According to Gibbons and Hawking \cite{Gibbons:1976ue}, $S_{\text{dS}} = \log \mathcal{Z}_{\text{grav}}$ calculates the quantum corrected entropy of the dS$_3$ horizon. To leading order in the large $\gamma$ expansion, $S_{\text{dS}} = 2\pi \gamma = \pi \ell/2 G$. The one-loop correction is given by the analytic continuation of the expression at integer level:
\begin{equation}\label{oneloopgrav}
e^{S^{(1)}_{\text{dS}}} = \left(\frac{1}{\text{vol} \, SU(2)} \right)^2 \times \left| \frac{4\pi^2}{\kappa+ i \gamma + 2} \right|^3~.
\end{equation}
This is the natural non-Abelian extension of (\ref{complexS3}). It is challenging to reproduce the leading contribution  $S_{\text{dS}}$ directly from the gravitational perspective (attempts include \cite{Maldacena:1998ih,Banados:1998tb,Govindarajan:2002ry}). In line with our discussion so far, we now consider the possibility of an edge-mode theory which may reproduce the subleading corrections to $S_{\text{dS}}$ from a Lorentzian perspective.

\subsection{Lorentzian signature}

We now turn to the Lorentzian theory
\begin{equation}\label{3dgravL}
S_L[\mathcal{A}_\mu] = \frac{k + i \lambda}{8\pi}  \int_{\mathcal{M}} d^3 x  \, \varepsilon^{\mu\nu\rho} \, \text{Tr} \left(\mathcal{A}_\mu \partial_\nu \mathcal{A}_\rho + \frac{2}{3}\mathcal{A}_\mu \mathcal{A}_\nu \mathcal{A}_\rho \right) + \text{c.c.}~,
\end{equation} 
where $\mathcal{A}_\mu$ is an $SL(2,\mathbb{C})$ gauge field related to the vielbein and spin connection as
\begin{equation} 
\mathcal{A}_\mu = \left(\omega^i_\mu  + i  e^i_\mu\right)T_i~, \quad \bar{\mathcal{A}}_\mu =\left(\omega^i_\mu  - i  e^i_\mu\right)T_i~,
\end{equation}
where $(T_1,T_2,T_3) = \tfrac{1}{2}(i\sigma_2, \sigma_1, \sigma_3)$ are the real generators of $SL(2,\mathbb{R})$ obeying $\text{Tr}(T_i T_j) = \tfrac{1}{2} \eta_{ij}$ and $[T_i,T_j] = \varepsilon_{ijk}T^k$, with $T^i = \eta^{ij}T_j$. We further have that $k\in\mathbb{Z}^+$ and $\lambda\in\mathbb{R}$. 

As for the Abelian theory with $U_{\mathbb{C}}(1)$ gauge group, we would like to understand the edge-mode theory. We must pick a boundary condition for the $SL(2,\mathbb{C})$ gauge field. Any choice will break the full diffeomorphism group, for the same reason that boundary conditions break the gauge-symmetries of the Chern-Simons theory. Various proposals for boundary conditions have appeared in the literature \cite{Carlip:1994gy,Banados:1996ad,Arcioni:2002vv}. Our interest is to understand the diffeomorphism invariant one-loop correction $S^{(1)}_{\text{dS}}$ (\ref{oneloopgrav}), and more generally the subleading corrections to the three-level de Sitter entropy $S_{\text{dS}}$ encoded in (\ref{subleadingterms}), from a Lorentzian perspective. In line with our previous discussions, we choose boundary conditions (\ref{gauge}), namely
\begin{equation}
\left( \mathcal{A}_t - \upsilon \mathcal{A}_\varphi \right) |_{\partial \mathcal{M}} = 0~, 
\end{equation}
with $\upsilon \in \mathbb{C}$. Going through the same steps, the Lorentzian edge-mode theory is then described by a chiral $SL(2,\mathbb{C})$ WZW theory at level $k + i \lambda$ \cite{Maldacena:1998ih,Witten:1989ip,Arcioni:2002vv}
\begin{equation}\label{WZWsl2C}
S_{\text{edge}}[g] = \frac{k + i \lambda}{8\pi} \, \text{Tr} \int dt d\varphi  \left( g  \partial_\varphi  g^{-1} \right) \left( g \left(\partial_t - \upsilon \partial_\varphi \right) g^{-1} \right)  + \pi (k + i \lambda) S^{\text{WZ}}[g] + \text{c.c.} ~,
\end{equation}
where $g(t,\varphi)$ is an $SL(2,\mathbb{C})$ valued function, and 
\begin{equation}
S^{\text{WZ}}[g] = \frac{1}{24 \pi^2} \int_{\mathcal{B}} d^3 y \, \varepsilon^{\mu \nu \rho} \, \text{Tr}\left(g^{-1} \partial_\mu g g^{-1} \partial_\nu g  g^{-1} \partial_\rho g\right) ~,
\end{equation}
where $\mathcal{B}$ is a three-manifold whose boundary is the $t\varphi$-cylinder.

As for the Abelian Lorentzian model, the Hamiltonian stemming from (\ref{WZWsl2C}) is real but unbounded from below. The theory satisfies an $SL(2,\mathbb{C})$ current algebra given by (see for instance \cite{Maldacena:1998ih})
\begin{equation}
\begin{split}
	 \left[ J^a_{n}, J^b_{m} \right] &= \tfrac{1}{2}(k + i \lambda ) n \delta_{ab} \delta_{n + m, 0} -  i f_{abc} \, J^c_{n+m}, \\
	 \left[ \tilde{J}^{a}_{n} , \tilde{J}^{b}_{m} \right] &= \tfrac{1}{2} (k - i \lambda ) n \delta_{ab} \delta_{n + m, 0} - i f_{abc}\, \tilde{J}^{c}_{n+m} ~,
\end{split}
\end{equation}
where the $SL(2,\mathbb{C})$ structure constants are $f_{abc} = \varepsilon_{abc}$, where $\varepsilon_{abc}$ is the Levi-Civita symbol. Using the above currents, the Sugawara construction yields a Virasoro algebra with central charge
\begin{equation}
c_{{SL}(2,\mathbb{C})} = \frac{3(k+i\lambda)}{(k+i\lambda+2)} +\frac{3(k-i\lambda)}{(k-i\lambda+2)} = \frac{6 \left(k (k+2) + \lambda ^2 \right)}{(k+2)^2+\lambda ^2}~. 
\end{equation}
Some details of the derivation for $c_{{SL}(2,\mathbb{C})}$ are provided in appendix \ref{centralcharge}. Notice that in the semiclassical limit $k\to\infty$, we have that $c_{{SL}(2,\mathbb{C})} \approx 6$ which is the number of field theoretic degrees of freedom in (\ref{WZWsl2C}). Interestingly, upon analytically continuing $k = -2$, the central charge is exactly $c_{{SL}(2,\mathbb{C})} = 6$, which might suggest that the theory is very simple at this point. Moreover, we see that $c_{\text{WZW}}$ is given by complexifying the level $k$ for the central charge $c_{{SU(2)}} = 3k/(k+2)$ of the $SU(2)$ WZW model. It also appears that the vacuum ($s=0$) $\widehat{\mathfrak{su}}(2)_k$ character (\ref{su2char}) can be analytically continued to complex level, giving the thermal partition function of this theory. 

As for the case of the Lorentzian complexified $U(1)$ model studied in section \ref{complexCSL}, upon continuing the Lorentzian edge-mode theory (\ref{3dgravL}) to Euclidean signature, one can consider complexifying the $SL(2,\mathbb{C})$ contour leading to that of a WZW model with two copies of $SU(2)$ at complex level. In particular, to render the path-integral better defined it is natural to pursue the same avenue as in section \ref{complexL}. Upon complexifying the contour of $g$ to one where it is valued in $SU(2)\times SU(2)$ we can make contact with (\ref{oneloopgrav}) much like (\ref{nonabelianedge}) connects to the three-sphere partition function of $SU(2)$ Chern-Simons theory (\ref{zsu2s3}). A detailed treatment of the putative gravitational edge-mode theory (\ref{WZWsl2C}) and its thermodynamic properties is left for future work. 

As a final remark, we note that timelike Wilson lines piercing the origin of the spatial disk carry unitary irreducible representations of $SL(2,\mathbb{C})$ which when non-trivial, contain an infinite number of states.\footnote{See \cite{Castro:2020smu} for a related discussion.} This is another indication of the non-compactness of $SL(2,\mathbb{C})$. It would be interesting to generalise the discussion in section \ref{Comments on the non-Abelian case} to this case.

\section{AdS$_4$/CFT$_3$ and $S^3$}\label{adscft}

In this final section we would like to consider the AdS$_4$/CFT$_3$ correspondence for an AdS$_4$ spacetime whose asymptotic boundary is given by a Euclidean or Lorentzian three-dimensional de Sitter space. This allows us to explore the thermodynamic properties of strongly coupled conformal matter theories on a fixed de Sitter background and their geometrisation in the bulk of AdS$_4$.

\begin{figure}[!h]
	\centering
	\includegraphics[width=0.4\linewidth]{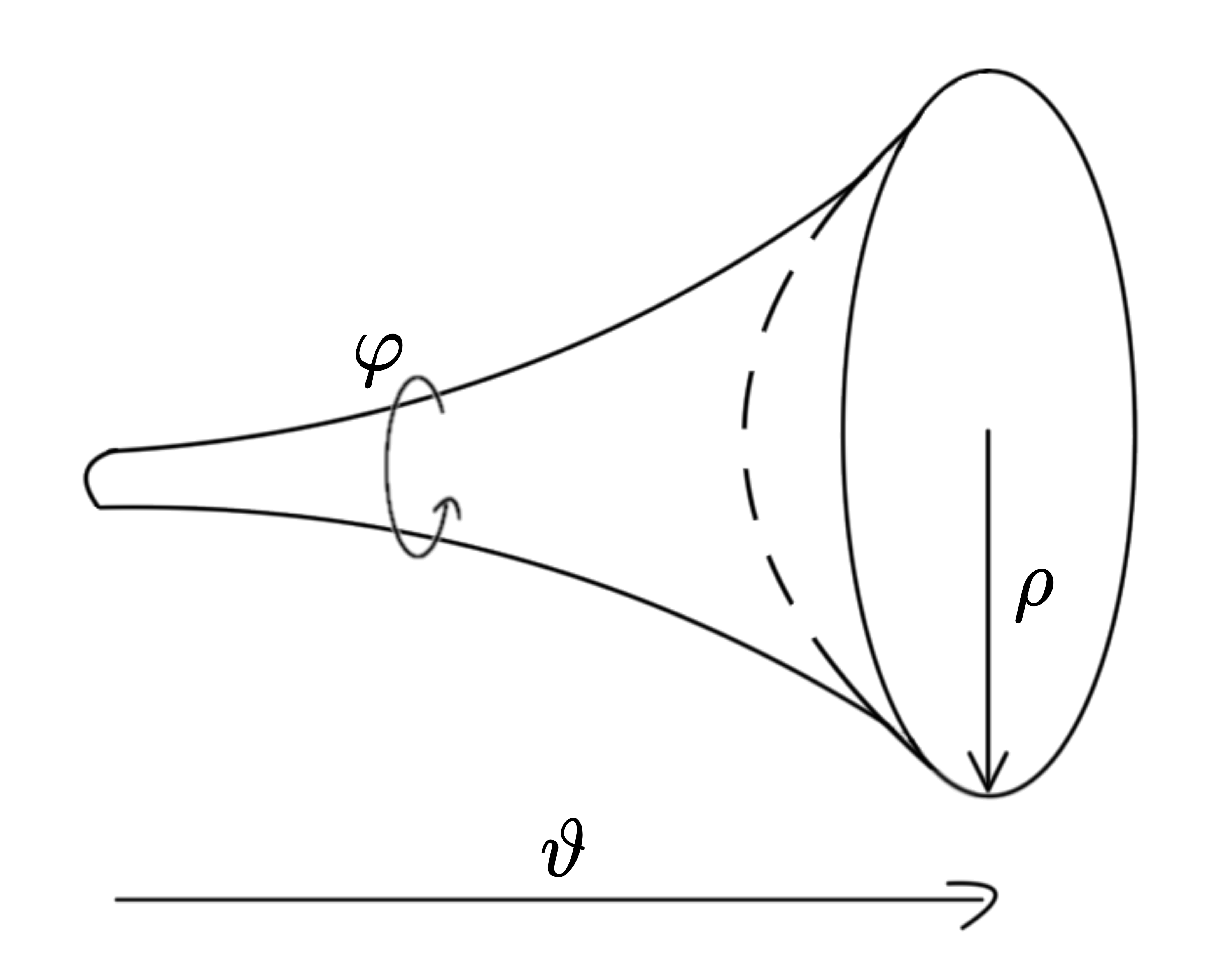}
	\caption{A constant $t$ slice of the geometry (\ref{LAdS}). The space is foliated by static patch hemispheres with exponentially increasing radius in the $\vartheta$ direction and the horizon at $\rho = \pi/2$ is the surface of this shape. }
	\label{fig:AdS4Spslicing}
\end{figure}

The metric of Euclidean AdS$_4$ is given by
\begin{equation}\label{EAdS}
\frac{ds^2}{\ell^2_{A}} = d\vartheta^2 + \sinh^2\vartheta \, d\Omega^2_3~,
\end{equation}
with $\vartheta \in [0,\infty)$ and $d\Omega_3^2$ describing the round metric on a unit $S^3$. The four-dimensional cosmological constant is $\Lambda = -3/\ell_A^2$. The asymptotic boundary resides at $\vartheta = \infty$ and the induced metric at the boundary is the round metric on $S^3$. Recalling our previous discussion, we can consider a Lorentzian continuation of (\ref{EAdS}) to the following static spacetime
\begin{equation}\label{LAdS}
\frac{ds^2}{\ell^2_{A}} = d\vartheta^2 + \sinh^2\vartheta \left(-dt^2 \cos^2 \rho + d\rho^2 + \sin^2\rho \, d\varphi^2 \right)~.
\end{equation}
This is a static Lorentzian anti-de Sitter universe whose constant-$\vartheta$ surfaces are given by the static patch of dS$_3$. A constant-$t$ slice of this geometry is shown in figure \ref{fig:AdS4Spslicing}. The global geometry is given by replacing the three-dimensional static patch metric in (\ref{LAdS}) with the global one. The static geometry (\ref{LAdS}) has a horizon at $\rho = \pi/2$. The topology of the horizon is $S^1 \times \mathbb{R}^+$. The Bekenstein-Hawking entropy of the horizon is given by
\begin{equation}\label{SBH}
S_{\text{BH}} =  \frac{\pi \ell_A^2}{2\ell_{pl}^2}  \int_0^{-\log \varepsilon} d\vartheta \sinh \vartheta \approx  \frac{\pi \ell_A^2}{2\ell_{pl}^2}\left( \frac{1}{2 \varepsilon } - 1+ \ldots \right)~,
\end{equation}
where $\ell_{pl}$ is the four-dimensional Planck length (such that $G = \ell_{pl}^2$) and $\varepsilon$ is a small number cutting off $z = e^{-\vartheta}$. From the perspective of AdS/CFT the entropy $S_{\text{BH}}$ corresponds to the entropy across the dS$_3$ horizon of the dual CFT, and the $1/\varepsilon$ divergence corresponds to a local divergence in the CFT due to entanglement of modes localised on the $S^1$ horizon. 

It is useful to expand the geometry in small $z$
\begin{equation}
\frac{ds^2}{\ell^2_A} = \frac{dz^2}{z^2} + \frac{1}{4} \left( \frac{1}{z^2} - 2 + z^2  \right) d\Omega^2_3~.
\end{equation}
From the above expansion it follows that the solution has vanishing boundary Brown-York stress tensor. Since the energy vanishes, we have that the logarithm of the thermal partition function is given entirely by the entropy. The thermal partition function is given by the Euclidean gravity path integral with an $S^3$ boundary, such that
\begin{equation}
\log Z[\text{EAdS}_4] \overset{?}{=} S_{\text{BH}}~.
\end{equation}
In the semiclassical limit, we can calculate $\log Z[\text{EAdS}_4]$ by evaluating the on-shell Einstein action 
\begin{equation}
S_E[g_{ij}] = - \frac{1}{16\pi \ell_{pl}^2}\int_{\mathcal{M}} d^4x \sqrt{g} \left(R + \frac{6}{\ell_A^2} \right) - \frac{1}{8\pi \ell_{pl}^2}\int_{\partial\mathcal{M}} d^3x \sqrt{h} K~,
\end{equation}
on the solution (\ref{EAdS}). The second term is the Gibbons-Hawking boundary term, with $K$ the extrinsic curvature at the boundary $\partial \mathcal{M}$ and $h_{ij}$ is the induced metric at the boundary measured in units of $\varepsilon$:
\begin{equation}
ds_{\text{bdy}}^2 = \frac{1}{4} d\Omega_3^2~.
\end{equation}
In general we can also add boundary terms which are built locally from the boundary metric $h_{ij}$. These will not affect the bulk equations of motion. Using a similar regularisation prescription as in (\ref{SBH}), we find
\begin{equation}
S_E = \frac{\pi \ell_A^2}{2\ell_{pl}^2}  + \frac{a_0}{\varepsilon^3} \int_{S^3} d^3 x \sqrt{h} + \frac{a_1}{\varepsilon} \int_{S^3} d^3 x \sqrt{h} R[h]~.
\end{equation}
The coefficients $a_0$ and $a_1$ depend on our choice of terms localised at the boundary. In the dual quantum field theory these are related to ultraviolet divergences (\ref{divergence}) renormalising the boundary cosmological constant and Newton constant. If our theory is supersymmetric, we expect that $a_0 = 0$. Thus, in the semiclassical limit we have
\begin{equation}\label{logZAdS}
\log Z[\text{EAdS}_4] = - \frac{\pi \ell_A^2}{2\ell_{pl}^2} - \frac{6 \pi^2 a_1}{\varepsilon}~.
\end{equation}
We see that the finite term indeed matches (\ref{SBH}). If we further tune $a_1 = -\ell_A^2/24 \pi \ell_{pl}^2$ we can also match the linearly divergent term. 

For ABJM theory with $SU(N)_k\times SU(N)_{-k}$ gauge group, the partition function (\ref{logZAdS}) has been calculated exactly in \cite{Kapustin:2009kz} and matched to the bulk in \cite{Drukker:2010nc}. The finite part, in terms of the ABJM data, reads
\begin{equation}
\log Z_{\text{ABJM}}[S^3] = -  N^2 \frac{\pi \sqrt{2}}{3} \lambda^{-1/2}~,
\end{equation}
where $\lambda = N/k$ is the 't Hooft parameter which is kept fixed and large in the large $N$ limit. The fact that $\log Z_{\text{ABJM}}[S^3]$ goes as $\sim N^2$ suggests that the theory is in a deconfined phase, reminiscent of the discussion in \cite{Witten:1998zw}. From our perspective, the calculations of \cite{Kapustin:2009kz, Drukker:2010nc} are a microscopic derivation of the Bekenstein-Hawking entropy of the horizon in (\ref{LAdS}). The Bekenstein-Hawking entropy (\ref{SBH}) predicts only linear ultraviolet divergences in the partition function. In fact, this is also true for the $S^3$ partition function calculated in \cite{Kapustin:2009kz}, since they regularise the theory in a way that preserves supersymmetry. The ultraviolet divergent piece of their calculation comes from the one-loop determinants of a transverse vector field and a gaugino on $S^3$. The absence of a cubic divergence is a result of the spectrum alone and is insensitive to any mass terms. 

Recalling (\ref{divergence}), the only other allowed divergence in a parity invariant three-dimensional quantum field theory is linear in $\ell^{-1}_{\text{uv}}$ and is indeed implicitly present in the localisation treatment of \cite{Kapustin:2009kz}. This is in agreement with (\ref{SBH}).  Interestingly, the calculations of \cite{Kapustin:2009kz,Drukker:2010nc} can be done for any value of the Chern-Simons level $k$ and rank $N$. They can be viewed as predicting string and loop corrections of the Bekenstein-Hawking entropy from the bulk perspective. In the perturbative limit, where $\lambda \ll 1$, it is found that 
\begin{equation}
\log Z_{\text{ABJM}}[S^3] \approx - N^2 \log \lambda^{-1} + \ldots~,
\end{equation}
which is in agreement with the pure Chern-Simons partition function result (\ref{CSpert}) in the perturbative limit.\footnote{The relation between topological entanglement entropy and black holes in AdS/CFT was also studied in \cite{McGough:2013gka}.}

One can also consider adding Wilson loops \cite{Kapustin:2009kz,Drukker:2010nc}. In the Euclidean picture, the simplest case is a Wilson loop that goes around the equator of the $S^3$.  To preserve supersymmetry, in addition to the gauge field, one also has additional matter fields on the loop. For instance in ABJM, the partition function endowed with a $1/2$-BPS preserving Wilson loop in the fundamental representation is given as follows
\begin{equation}\label{WL}
\frac{Z_\text{ABJM}[S^3;W_{\mathcal{C}}]}{Z_{\text{ABJM}}[S^3]} = \frac{e^{i\pi/2} }{2} e^{\pi\sqrt{2\lambda}}~,
\end{equation}
where the expression is given to leading order in the large $N$ limit at fixed and large 't Hooft coupling $\lambda = N/k$. From the bulk perspective, the above is computed by a string whose worldsheet intersects the $S^3$ boundary at the equatorial $S^1$. The phase is due to a type IIA B-field. We can  Wick rotate to a Lorentzian picture (\ref{LAdS}), where the equatorial $S^1$ at $\rho=0$ is now parameterised by the Lorentzian coordinate $t$. The worldsheet now intersects the boundary along $t$ at $\rho = 0$ and goes all the way to the horizon. In the thermodouble field picture, we can continue the worldsheet across the horizon such that the worldsheet intersects the second boundary static patch. As in our discussion of pure Chern-Simons theory, we might view the partition function with the insertion of the Wilson loop as computing a part of the entanglement entropy between the two static patches in the presence of an insertion at the origin of the two spatial disks. From the bulk perspective this is the contribution to the Bekenstein-Hawking entropy (\ref{SBH}) due to a worldsheet crossing the bulk horizon.

\section*{Acknowledgements}

It is a great pleasure to acknowledge Tarek Anous, Frederik Denef, Nadav Drukker, Diego Hofman, and Beatrix M\"uhlmann for insightful discussions. We especially thank Beatrix M\"uhlmann and Tarek Anous for a thorough reading and comments on the draft.  D.A. is funded by the Royal Society under the grant The Atoms of a deSitter Universe. E.H. is funded by an STFC studentship ``Aspects of black hole and cosmological horizons". 

\appendix
\section{Chern-Simons determinant on $S^3$}\label{S3det}

In this appendix we consider the determinant that stems from the quadratic part of Chern-Simons theory on $S^3$. We first recall that the canonical mass dimension of $A_\mu(x)$ is zero, as is the mass dimension for the gauge parameter $\alpha(x)$. We define the normalisation of our path integration measures
\begin{eqnarray}\label{norm}
1 &\equiv& \int \mathcal{D} \alpha \, e^{-\ell_{\text{uv}}^{-3}\int d^3 x \sqrt{g} \alpha(x)^2/2}~, \\ 
1 &\equiv& \int \mathcal{D} A^T_\mu \, e^{-\ell_{\text{uv}}^{-1}\int d^3 x \sqrt{g} g^{\mu\nu} A^T_\nu(x) A^T_\mu(x)/2}~,  \\
1 &\equiv& \int \mathcal{D} \bar{c}  \mathcal{D} c \, e^{\ell_{\text{uv}}^{-2}\int d^3 x \sqrt{g}   \bar{c}(x) {c}(x)}~, 
\end{eqnarray}
where $\nabla^\mu A^T_\mu(x) = 0$, $\alpha(x)$ is a real scalar, while $\bar{c}(x)$ and $c(x)$ are Grassmann valued fields. To render the exponents dimensionless, we multiply by the appropriate powers of the UV cutoff length scale $\ell_{\text{uv}} = 1/\Lambda_{\text{uv}}$. In fixing the above normalisation, we fix any ultraviolet ambiguities stemming from field rescalings. The fields can be expanded in a complete basis of eigenfunctions of the Laplacian on the round three-sphere. For instance, 
\begin{equation}
\alpha(x) = \sum_l \alpha_l \phi_l(x)~, \quad\quad -\nabla^2 \phi_l = \xi_l \phi_l~, \quad\quad \int d^3 x \sqrt{g} \phi_l(x) \phi_{l'}(x) = \delta_{l l'}~,
\end{equation}
such that mass dimension of $\alpha_l$ is $-3/2$. Following similar steps for the other fields, we obtain the following path integration measures
\begin{equation}
\mathcal{D} \alpha \equiv  \prod_{l} {\Lambda_{\text{uv}}^{3/2}} \,  \frac{d\alpha_l}{{\sqrt{2\pi}}}~, \quad \mathcal{D} \bar{c} \mathcal{D} c \equiv  {\prod}'_{l} \,{\Lambda_{\text{uv}}^{-2}} d\bar{c}_l d c_l~,  \quad \mathcal{D} A^T_\mu \equiv \prod_l \Lambda_{\text{uv}}^{1/2} \frac{dA_l}{\sqrt{2\pi}}~.
\end{equation}
With the above definitions of the path integration measures, and the Chern-Simons action normalised as
\begin{equation}
S_{\text{CS}} = \frac{1}{2} \int d^3x  \, \varepsilon^{\mu\nu\rho} A_\mu \partial_\nu A_\rho~,
\end{equation} 
we can proceed via the Fadeev-Popov gauge fixing procedure. We take the metric on $S^3$ to be
\begin{equation}
ds^2 = \ell^2 \left(d\theta^2 + \sin^2\theta d\Omega_2^2 \right)~,
\end{equation}
with volume $\text{vol}\, S^3 = 2\pi^2 \ell^3$. Working in the Lorenz gauge $\nabla_\mu A^\mu = 0$, one finds 
\begin{equation}\label{fulldet}
|Z_{U(1)_k}[S^3] | = \sqrt{\frac{2\pi}{k}}  \times \frac{\sqrt{2\pi}(\text{vol} \, S^3 \Lambda_{\text{uv}}^3)^{-1/2}}{\text{vol} \, U(1)} \times \sqrt{\frac{{\det}'[ -\nabla^2 / \Lambda_{\text{uv}}^2]}{{\det[ L L^\dag /\Lambda_{\text{uv}}^2]}^{1/2}}}~,
\end{equation}
where $\text{vol} \, U(1) = 2\pi$. The term $\sqrt{2\pi}(\text{vol} \, S^3 \Lambda_{\text{uv}}^3)^{-1/2}/\text{vol} \, U(1)$ follows from the zero mode contribution $d\alpha_0$. It originates from the residual part of the gauge group volume that is not cancelled by the Lorenz gauge.  

After the dust settles, we are left to evaluate the following ratio of functional determinants
\begin{equation}
r^2 =  \frac{{\det}'\left(-\nabla^2 / \Lambda_{\text{uv}}^2\right)}{\sqrt{\det L L^\dag /\Lambda_{\text{uv}}^2}}~,
\end{equation}
where the Laplacian $-\nabla^2$ acts on scalar functions on $S^3$ and $L L^\dag$ acts on transverse vector fields on $S^3$. The prime indicates we are dropping the zero mode of $-\nabla^2$. The respective spectra are well known \cite{Rubin:1984tc}. For the scalar Laplacian we have
\begin{equation}
\lambda_n = \frac{1}{\ell^2}\times n(n+2)~, \quad\quad d_n = (n+1)^2~, \quad\quad n = 0,1,\ldots
\end{equation}
For the $L L^\dag$ operator, the eigenvalues and degeneracies are given by
\begin{equation}
\lambda_n = \frac{1}{\ell^2}\times (n+1)^2~, \quad\quad d_n = 2n(n+2)~, \quad\quad n = 1,2,\ldots
\end{equation}
Using a heat kernel regularisation,\footnote{For a $\zeta$-function regularisation scheme, see \cite{Fliss:2017wop}.} we have that
\begin{equation}
-\log {\det}'\left(-\frac{\nabla^2}{\Lambda_{\text{uv}}^2}\right) = \sum_{n=1}^\infty (n+1)^2 \, \int_{0}^\infty \frac{d\tau}{\tau} e^{-\frac{\varepsilon^2}{4\tau}}  e^{-\tau n(n+2)}~,
\end{equation}
where the dimensionless cutoff parameter is taken to be $\varepsilon \equiv 2 e^{-\gamma}/\ell\Lambda_{\text{uv}}$ with $\gamma$ being the Euler constant. It is convenient to rewrite the above in the following form
\begin{equation}
-\log {\det}'\left(-\frac{\nabla^2}{\Lambda_{\text{uv}}^2}\right) = \int_{\mathcal{C}} \frac{du}{\sqrt{u^2 + \varepsilon^2}} e^{-i \sqrt{u^2+\varepsilon^2}} \left( \frac{1+e^{i u}}{1-e^{i u}} \frac{e^{i u}}{(1-e^{i u})^2} - e^{i u} \right)~.
\end{equation}
The contour $\mathcal{C} = \mathbb{R}+i\delta$ is parallel to the real axis but has a small positive imaginary part $0<\delta < \varepsilon$. We can deform the contour to go down the branch cut from the left and up from the right. We find
\begin{equation}
-\log {\det}'\left(-\frac{\nabla^2}{\Lambda_{\text{uv}}^2}\right) = \int_{\varepsilon}^\infty \frac{dt}{\sqrt{t^2 - \varepsilon^2}} \left( \frac{1+e^{-t}}{1-e^{-t}} \frac{e^{-t}}{(1-e^{-t})^2} - e^{-t} \right) \left( e^{\sqrt{t^2-\varepsilon^2}} + e^{-\sqrt{t^2-\varepsilon^2}} \right)~.
\end{equation}
Similar considerations lead to 
\begin{equation}
-\frac{1}{2} \log {\det} \, L L^\dag = 2 \int_{\varepsilon}^\infty \frac{dt}{\sqrt{t^2 -\varepsilon^2}} \frac{3 e^{-2 t} -e^{-3 t} }{\left(1-e^{-t}\right)^3}~.
\end{equation}
The ultraviolet behaviour is encoded in the small-$t$ regime of the integrands. Setting $\varepsilon = 0$ in the integrand and performing a small-$t$ expansion we find that there is no $\sim 1/\varepsilon^3$ divergence, and moreover that the leading divergence goes as $\sim 1/\varepsilon$. There is also a logarithmic $\sim \log \varepsilon$ divergence. Since we are in odd dimensions there should be no logarithmic divergences contributing to $\log Z[S^3]$. Recalling the expression (\ref{fulldet}) we see that this is indeed the case once we take into account the overall factor $\ell_{\text{uv}}^{-3/2}$. Evaluating the $t$-integral in the heat kernel regularisation in the small-$\varepsilon$ limit leads to 
\begin{equation}
\log r^2 = - \frac{3 \pi }{2 \varepsilon } -3 \log \frac{e^{\gamma} \varepsilon}{2} +\log 2 \pi ^2 = - \frac{3 \pi }{2 \varepsilon } + 3 \log \ell\Lambda_{\text{uv}} +\log 2 \pi ^2 ~.
\end{equation} 
Combining the above with (\ref{fulldet}) and recalling that $\text{vol}\, S^3 = 2\pi^2 \ell^3$, we find
\begin{equation}\label{finaldet}
|Z_{U(1)_k}(S^3)| = \sqrt{\frac{1}{k}} e^{-\frac{3\pi}{4\varepsilon}}~.
\end{equation}
A similar picture holds for Chern-Simons theories with more general gauge group in the perturbative, large $k$, regime. 
The coefficient of the $1/\varepsilon$ term in (\ref{finaldet}) can be tuned by adding a local background dependent term
\begin{equation}
S_{\text{b}} = \Lambda_b \int d^3 x \sqrt{g} R = 12\pi^2 \ell \Lambda_b~,
\end{equation}
where we recall that $R = 6/\ell^2$. For the particular choice $\Lambda_b =  -{e^{\gamma} \Lambda_{\text{uv}} }/{32 \pi }$ we can set the divergent term to zero. 

As a final remark, we note that the choice of normalisations (\ref{norm}) serve another physical purpose, namely to set the partition function of the theory on $S^1 \times S^2$ equal to unity (up to $1/\varepsilon$ divergences which can be absorbed into local counterterms). This agrees well with the fact that Chern Simons theory quantised on a spatial $S^2$ has a unique state with vanishing energy. To compute the partition function in this case, one must take into account the non-trivial moduli space of flat connections due to the presence of a non-contractible cycle \cite{Rozansky:1993rt}. The volume of this moduli space is crucial to cancel the $k$ dependence of the partition function. 

\section{Euclidean path-integral of Abelian edge-mode theory}\label{Zthermal}

In this appendix we provide a derivation of the thermal partition function (\ref{zedge}) from the Euclidean path-integral. We must Wick rotate the edge-mode theory (\ref{edge}) to Euclidean signature by taking $t = -i \tau$ and imposing the thermal periodicity condition $\tau \sim \tau + \beta$. Notice that although the original Chern-Simons theory, being topological, is insensitive to the signature of spacetime, the edge-mode theory can sense it. The reason for this is that from the gauge theory perspective we should also continue $A_t = i A_\tau$ when continuing $t = -i \tau$. This would violate the reality conditions of the gauge-fixing condition (\ref{gauge}) unless we also continue $\upsilon = - i \upsilon_E$. If we also analytically continue $\upsilon$, the edge-mode theory would remain unchanged, and hence unaware of the underlying signature. For the purposes of computing the thermal partition function, we keep $\upsilon$ unchanged upon continuation to Euclidean signature.

It is convenient to expand the non-winding sector in a Fourier basis
\begin{equation}
\zeta(\tau,\varphi) = \frac{1}{\sqrt{2\pi}} \sum_{(m,n) \in \mathbb{Z}^2} e^{2\pi m i \tau/\beta +  i n \varphi} \zeta_{m,n}~,
\end{equation}
with $\bar{\zeta}_{m,n} = \zeta_{-m,-n}$. Performing the Gaussian path-integral over the non-winding sector, we are led to evaluate
\begin{equation}
Z_{\text{non-winding}}[\beta] = \mathcal{N} \prod_{m, n=1}^\infty \frac{1}{1 + (\beta\upsilon n/2\pi m)^2}~.
\end{equation}
We will be rather cavalier about the overall normalisation $\mathcal{N}$ as we can determine it by imposing that there is a unique vacuum state and hence $\lim_{\beta\to\infty} \left(1- \beta\partial_\beta \right) \log Z_{\text{non-winding}}[\beta] = 0$. Using the infinite product representation of $\sinh z$, we can express the above as
\begin{equation}
Z_{\text{non-winding}}[\beta] = e^{-\beta\upsilon \epsilon_0} \times e^{ \beta \upsilon/24} \prod_{n=1}^\infty \frac{1}{1- e^{-\beta\upsilon n}} = \frac{e^{-\beta \upsilon \epsilon_0}}{\eta(q)}~,
\end{equation}
where we have absorbed any divergences of the overall energy scale into $\epsilon_0$ and $q = e^{-\upsilon\beta}$. If we further incorporate the winding mode sector wrapping the $\varphi$ cycle, we get an additional contribution given by 
\begin{equation}
Z_{\text{winding}}[\beta] = \sum_{n\in\mathbb{Z}} e^{-k \upsilon \beta n^2/2} = \vartheta_3(0,q^{k/2})~.
\end{equation}
Combining the two, we find agreement with (\ref{zedge}).

\section{Lorentzian edge-mode Hamiltonian}\label{LedgeApp}

\begin{figure}[H]
	\centering
	\includegraphics[width=0.5\linewidth]{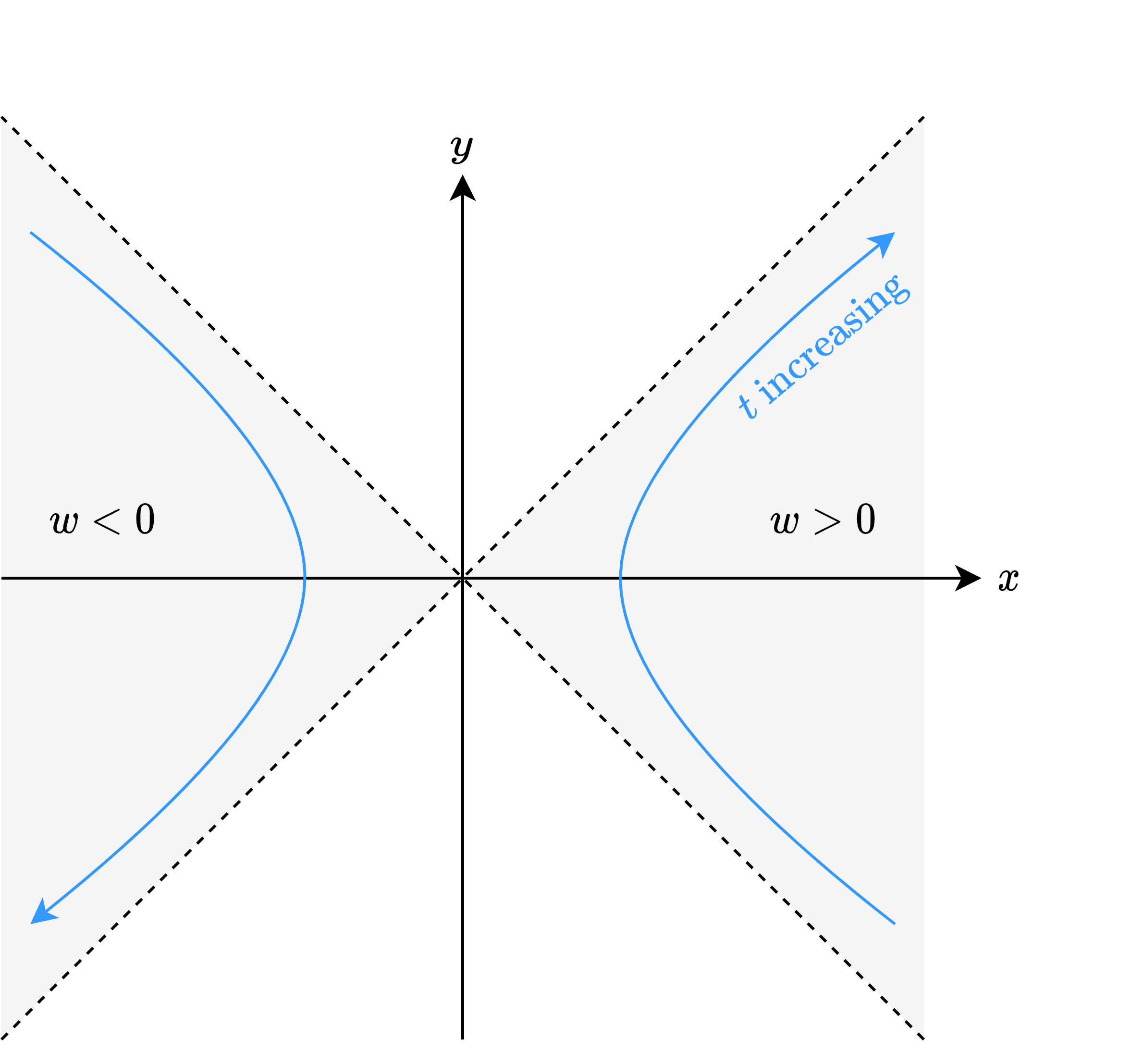}
	\caption{Plot showing the shaded regions $|x| > |y|$, where the Hamiltonian (\ref{xgryH}) is valid. The blue arrows indicate increasing $t = \tanh^{-1}(y/x)$. By gluing this solution along $|x| = |y|$ with the one in the $|y|>|x|$ region, we arrive at the solution (\ref{Loreenergies}).}
	\label{fig:xygraphSdS}
\end{figure}

In this appendix we show how the Lorentzian edge-mode Hamiltonian reduces to that of a conformal quantum mechanics. We start with the Hamiltonian in the form (\ref{LedgeH}). Since the modes carrying fixed momentum $n$ on the circle decouple, it is sufficient to study the Hamiltonian of a single mode number $n$: 
\begin{equation}
\hat{\mathcal{Q}}^{(n)}_t = \frac{\sqrt{k^2+\lambda^2}}{8\pi} n^2 \left( \upsilon_{re} \left( \hat{A}_n^+ \hat{A}_n^- - \hat{B}_n^+ \hat{B}_n^- \right) + i  \upsilon_{im} \left( \hat{A}_n^+ \hat{B}_n^+ - \hat{A}_n^- \hat{B}_n^-  \right) \right)~.
\end{equation}
where $\hat{\mathcal{Q}}_t  = \tfrac{1}{2} \sum_n \hat{\mathcal{Q}}^{(n)}_t$. We represent the raising and lowering operators as 
\begin{equation}
\hat{A}^\pm_n = \left( \frac{4\pi}{n\sqrt{k^2 + \lambda^2}}  \right)^{1/2} \left( x_n \pm \frac{d}{dx_n} \right)~, \quad \hat{B}^\pm_n =  \left( \frac{4\pi}{n\sqrt{k^2 + \lambda^2}}  \right)^{1/2} \left( y_n \pm \frac{d}{dy_n} \right)~.
\end{equation}
In these new coordinates, the Hamiltonian is 
\begin{equation}
\hat{\mathcal{Q}}^{(n)}_t = \frac{n}{2} \left[ \upsilon_{re} \left( - \left(\frac{d}{dx_n}\right)^2 + \left(\frac{d}{dy_n}\right)^2 + x_n^2 - y_n^2\right) + 2i  \upsilon_{im} \left( x_n \frac{d}{dy_n}  + y_n \frac{d}{dx_n} \right) \right]~.
\end{equation}
We then make the following change of coordinates: 
\begin{equation}
\begin{split} \label{HyperboilicCoordinates}
x_n = w_n \cosh t_n , \quad\quad  y_n = w_n \sinh t_n~,  \quad\quad (t_n,w_n) \in \mathbb{R}^2~,\\
\end{split}
\end{equation}
covering the region $|x|>|y|$, shown in figure \ref{fig:xygraphSdS}. In terms of $(w_n, t_n)$, our Hamiltonian is 
\begin{equation} \label{xgryH}
\hat{\mathcal{Q}}^{(n)}_t = \frac{n}{2} \left[ \upsilon_{re} \left( - \frac{d^2}{dw_n^2 } - \frac{1}{w_n} \frac{d}{dw_n} + \frac{1}{w_n^2} \frac{d^2}{dt_n^2}+ w_n^2\right) + 2i  \upsilon_{im} \frac{d}{dt_n} \right]~.
\end{equation}
For $|x|<|y|$ the Hamiltonian is given by
\begin{equation}
\hat{\mathcal{Q}}^{(n)}_t = - \frac{n}{2} \left[ \upsilon_{re} \left( - \frac{d^2}{dw_n^2 } - \frac{1}{w_n} \frac{d}{dw_n} + \frac{1}{w_n^2} \frac{d^2}{dt_n^2}+ w_n^2\right) + 2i  \upsilon_{im} \frac{d}{dt_n} \right]~.
\end{equation}
We can then find the Schr\"{o}dinger equation for each mode. By expanding the wavefunction in a Fourier basis as 
\begin{equation}
\Psi (w_n,t_n) = \int_{\mathbb{R}} \frac{dp_n}{2\pi} \,  \psi_n (w_n) e^{i p_n t_n}, \quad\quad p_n \in \mathbb{R}~,
\end{equation}
and taking $|x|>|y|$, we find
\begin{equation}
\frac{1}{2}\left( - \frac{d^2}{dw_n^2 } - \frac{1}{w_n} \frac{d }{dw_n} - \frac{p_n^2}{w_n^2} + w_n^2\right)\psi_n   = \frac{1}{\upsilon_{re} }\left(\frac{E_n}{n}+   \upsilon_{im} p_n \right)\psi_n .
\end{equation} 
To get this in the Shr\"{o}dinger form, we then make the substitution $\psi_n \rightarrow \frac{1}{\sqrt{w_n}} \psi_n$, and this becomes 
\begin{equation} \label{ComformalQuantumMechaics}
\frac{1}{2}\left(- \frac{d^2}{dw_n^2 } +w_n^2 - \frac{1/4 + p_n^2}{w_n^2}\right) \psi_n = \frac{1}{\upsilon_{re} } \left( \frac{E_n}{n} +  \upsilon_{im} p_n\right) \psi_n ,
\end{equation}
which is the conformal quantum mechanics problem studied in \cite{de1976conformal, Anous:2020nxu}. We need to patch this solution to the one for $|y| > |x|$ which can be found by exchanging $x_n \leftrightarrow y_n$ in (\ref{HyperboilicCoordinates}) and following the same steps. The result is a change of sign in the left-hand side of (\ref{ComformalQuantumMechaics}). 

\section{$SL(2,\mathbb{C})$ WZW central charge}\label{centralcharge}
To calculate the current algebra and central charge of the chiral $SL(2,\mathbb{C})$ WZW model (\ref{WZWsl2C}), we first study the non-chiral WZW model and then take the anti-holomorphic sector to describe the current algebra of the chiral theory. This is because the chiral theory is invariant under 
\begin{equation}
	g(z,\bar{z}) \rightarrow g(z,\bar{z}) \bar{\Omega}^{-1}(\bar{z}), 
\end{equation}
where $(z,\bar{z})$ are coordinates on the complex plane. The $SL(2,\mathbb{C})$ WZW model is given by 
\begin{equation}
	 S = \frac{k + i \lambda}{8\pi} \int d^2x \, \text{Tr} \left( \partial^\mu g^{-1} \partial_\mu g \right) - \frac{i (k + i \lambda)}{12 \pi} \int d^3 y \, \varepsilon^{\mu \nu \rho} \, \text{Tr} \left( g^{-1} \partial_\mu g g^{-1} \partial_\nu g g^{-1} \partial_\rho g\right) + \text{c.c.} ~,
\end{equation}
where $g \in SL(2,\mathbb{C})$. By computing the equations of motion for this action, we find the following conserved currents 
\begin{equation}
	\begin{split}
		J_z = - \tfrac{1}{2} (k + i \lambda ) \partial_z g g^{-1} ~, &\qquad  J_{\bar{z}} = \tfrac{1}{2} (k + i \lambda ) g^{-1} \partial_{\bar{z}} g ~,\\
		\tilde{J}_{\bar{z}} = - \tfrac{1}{2} (k - i \lambda ) \partial_{\bar{z}} \bar{g} \bar{g}^{-1}~ , &\qquad  \tilde{J}_{z} = \tfrac{1}{2} (k - i \lambda ) \bar{g}^{-1} \partial_{z} \bar{g} ~,
	\end{split}
\end{equation}
Here the subscripts label which currents are holomorphic/ anti-holomorphic. We can calculate the current algebra (using the procedure described for instance in \cite{difrancesco}) and find
\begin{equation}
	\begin{split}
		\left[ J^a_{z,n}, J^b_{z,m} \right] &= \tfrac{1}{2}(k + i \lambda ) n \delta_{ab} \delta_{n + m, 0} + \sum_c i f_{abc} J^c_{z,n+m} ~,\\
		\left[ J^a_{\bar{z},n}, J^b_{\bar{z},m} \right] &= \tfrac{1}{2} (k + i \lambda ) n \delta_{ab} \delta_{n + m, 0} - \sum_c i f_{abc} J^c_{\bar{z},n+m}~, \\
		\left[ \tilde{J}^{a}_{\bar{z},n}, \tilde{J}^{b}_{\bar{z},m} \right] &= \tfrac{1}{2} (k - i \lambda ) n \delta_{ab} \delta_{n + m, 0} - \sum_c i f_{abc} \tilde{J}^{c}_{\bar{z},n+m} ~,\\
		\left[ \tilde{J}^{a}_{z,n}, \tilde{J}^{b}_{z,m} \right] &= \tfrac{1}{2} (k - i \lambda ) n \delta_{ab} \delta_{n + m, 0} + \sum_c i f_{abc} \tilde{J}^{c}_{z,n+m} ~.
	\end{split}
\end{equation}
The anti-holomorphic sector is then given by the second and third of these. Following the Sugawara construction, we find the anti-holomorphic energy-momentum tensor
\begin{equation}
	\bar{T}(\bar{z}) = \frac{1}{k + i \lambda +2} \sum_a (J^a_{\bar{z}} J^a_{\bar{z}} ) (\bar{z}) + \frac{1}{k - i \lambda +2} \sum_a (\tilde{J}^a_{\bar{z}} \tilde{J}^a_{\bar{z}} ) ({\bar{z}}) ~,
\end{equation}
where we have used $f_{abc} f_{cbd} = \epsilon_{abc} \epsilon_{cbd} = 2 \delta_{ad}$. From the OPE
\begin{equation}
	\bar{T}(z) \bar{T}(w) =  \left(\frac{3(k+i \lambda)}{2(k + i \lambda + 2)} + \frac{3(k-i \lambda) }{2(k - i \lambda + 2)} \right)  \frac{1}{(\bar{z}-\bar{w})^4} + \frac{2\bar{T}(\bar{w})}{(\bar{z}-\bar{w})^2} + \frac{\partial \bar{T}(\bar{w})}{\bar{z}-\bar{w}}~,
\end{equation}
we can find the central charge 
\begin{equation}
c_{{SL}(2,\mathbb{C})} = \frac{3(k+i\lambda)}{(k+i\lambda+2)} +\frac{3(k-i\lambda)}{(k-i\lambda+2)} = \frac{6 \left(k (k+2) + \lambda ^2 \right)}{(k+2)^2+\lambda ^2}~. 
\end{equation}

\bibliographystyle{unsrturl}
\bibliography{SoftdeSitterbibliography}{}

\end{document}